\documentclass[%
reprint,
superscriptaddress,
 amsmath,amssymb,
 aps,
pra,
]{revtex4-1}

\usepackage{placeins}
\usepackage{siunitx}
\usepackage[table]{xcolor}
\usepackage{xcolor}
\usepackage{graphicx}
\usepackage{dcolumn}
\usepackage{bm}
\usepackage{hyperref}
\usepackage{xfrac}
\usepackage{gensymb}

\newcommand{\alex}[1]{\textbf{\bf\textcolor{orange}{#1}}}

\newcommand{\fMW}{\nu}
\newcommand{\fUp}{\fMW_{+1}}

\newcommand{\fDown}{\fMW_{-1}}

\newcommand{\Bbias}{B_0}
\newcommand{\mS}{m_\mathrm{s}}

\newcommand{\SplOne}{S_\mathrm{+1}}
\newcommand{\SminOne}{S_\mathrm{-1}}
\newcommand{\Ssum}{\bar{S}}
\newcommand{\Sdiff}{\Delta S}

\begin{document}

\preprint{APS/123-QED}

\title{Direct control of high magnetic fields for cold atom experiments based on NV centers}

\author{Alexander Hesse}
\affiliation{Kirchhoff-Institut f\"ur Physik, Im Neuenheimer Feld 227, 69120 Heidelberg}

\author{Kerim K\"oster}
\affiliation{Kirchhoff-Institut f\"ur Physik, Im Neuenheimer Feld 227, 69120 Heidelberg}

\author{Jakob Steiner}
\affiliation{3. Physikalisches Institut, Center for Applied Quantum Technologies, IQST , Pfaffenwaldring 57, 70569 Stuttgart}
\affiliation{Paul-Scherrer-Institute, 5323 Villigen, Switzerland}

\author{Julia Michl}
\affiliation{3. Physikalisches Institut, Center for Applied Quantum Technologies, IQST , Pfaffenwaldring 57, 70569 Stuttgart}

\author{Vadim Vorobyov}
\affiliation{3. Physikalisches Institut, Center for Applied Quantum Technologies, IQST , Pfaffenwaldring 57, 70569 Stuttgart}

\author{Durga Dasari}
\affiliation{3. Physikalisches Institut, Center for Applied Quantum Technologies, IQST , Pfaffenwaldring 57, 70569 Stuttgart}

\author{J\"org Wrachtrup}
\affiliation{3. Physikalisches Institut, Center for Applied Quantum Technologies, IQST , Pfaffenwaldring 57, 70569 Stuttgart}
\affiliation{Max Planck Institute for Solid State Research, Heisenbergstraße 1, 70569 Stuttgart, Germany}

\author{Fred Jendrzejewski}
\affiliation{Kirchhoff-Institut f\"ur Physik, Im Neuenheimer Feld 227, 69120 Heidelberg}

\date{November 9, 2020}

\begin{abstract}

In cold atomic gases the interactions between the atoms are directly controllable through external magnetic fields. The magnetic field control is typically performed indirectly by stabilizing the current through a pair of Helmholtz coils, which produce this large bias field. Here, we overcome the limitations of such an indirect control through a direct feedback scheme, which is based on nitrogen-vacancy centers acting as a magnetic field sensor. This allows us to measure and stabilize fields of \SI{4.66}{\milli \tesla} down to \SI{12}{\nano \tesla} RMS noise over the course of \SI{24}{\hour}, measured on a \SI{1}{\hertz} bandwidth. We achieve a control of better than $ \SI{1}{ppm}$ after 20 minutes of integration time, ensuring high long-term stability for experiments. This approach extends direct magnetic field control to strong magnetic fields, which could enable new precise quantum simulations in this regime.

\end{abstract}
\maketitle

\section{\label{sec:Int}Introduction}

The direct control of modest magnetic fields up to \SI{1}{\milli \tesla} enabled experiments with cold atoms on diverse topics, like spin squeezing close to a Feshbach resonance \cite{StrobelFisher, MuesselSqueezing}, the quantum simulation of scaling dynamics far from equilibrium \cite{MaxNature} or of a scalable building block for lattice gauge theories in atomic mixtures \cite{MilGauge}. In all these cases, the direct magnetic field control is based on active feedback from a fluxgate sensor, which limits this class of experiments to magnetic fields below \SI{1}{\milli \tesla}, the working range of fluxgate sensors \cite{Fluxgate_review}. Many phenomena, like droplet formation close to a heteronuclear Feshbach resonance \cite{DropletsLetitia} or spin changing collisions between fermionic and bosonic species \cite{Zache2018}, do however occur at much higher field strengths, reaching up to a few tens of \si{\milli \tesla}. Direct magnetic field control has so far not been possible in this regime due to a lack of suitable sensors, making precision experiments challenging.

This shows the need for alternative sensors able to measure large magnetic fields with high precision. Over the last years, several sensors based on quantum systems with the potential to close this gap have been tested, summarized in Table~\ref{Tab:Magnetometers}. Most prominently, SQUIDs can measure magnetic fields with sensitivities in the $\si{\sfrac{\femto \tesla}{\sqrt{\hertz}}}$ regime \cite{SQUID1, SQUID2}, with atomic gases reaching comparable or slightly better sensitivities at low fields \cite{Dang2010,Griffith2010, Mitchell_Mag, Magnetom_Oberth}. However, the large size of these systems as well as the demanding setup required for probing them makes them impractical as a sensor for an active magnetic field stabilization.

The nitrogen-vacancy (NV) center in diamond provides a versatile and compact magnetic field sensor, covering ranges up to several tens of \si{\tesla} with precisions below $\SI{1} {\sfrac{\pico \tesla}{\sqrt{\hertz}}}$ \cite{Subpico_Mag} for AC magnetic fields. For measurements which approach the DC regime, sensitivities of a few $\SI{10} {\sfrac{\pico \tesla}{\sqrt{\hertz}}}$ and below have been reported \cite{Barry2013, Walsworth_Vec, Clevenson2015,Fescenko2019}. Due to their small size, they are the perfect candidate for magnetic field stabilizations \cite{Glenn2018, Jaskula2019} past the range of fluxgate sensors, and allow the study of spin dynamics in regimes previously not experimentally accessible.

\begin{figure*}[t]
\includegraphics[width = \textwidth]{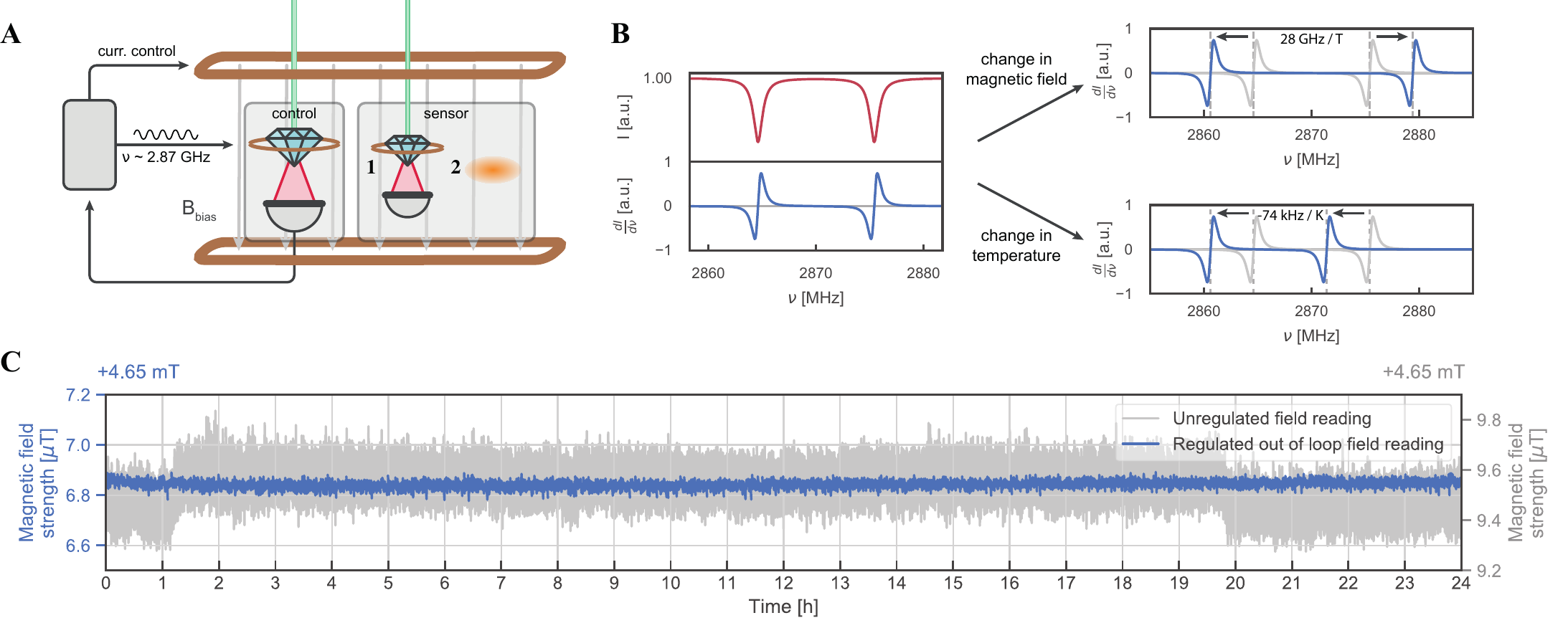}
\caption{\label{fig:total}\textbf{(A)} Basic experimental setup: The NV centers are positioned in a magnetic field generated by a pair of Helmholtz coils. They are excited by a green laser, and their red fluorescence is collected on a photodiode. This signal is fed into a control circuit, which also generates the microwave signal applied to the NV centers via a wire loop. From the fluorescence a magnetic field reading is extracted, and used to regulate the current through a second, low inductance coil pair. \textbf{(1)} The quality of the magnetic field stabilization is monitored on a second, independent setup placed close to the first sample. \textbf{(2)} In the future, this second NV sensor will be replaced by the cold quantum gas used in our experiments, such that it experiences stable magnetic fields. \textbf{(B)} Shift of the two transitions $\mS = 0 \rightarrow \mS = -1$ and $\mS = 0 \rightarrow \mS = 1$ with temperature and magnetic field, as visible in the fluorescence signal. Using a lock-in scheme we extract the derivative of the spectrum. \textbf{(C)} \SI{24}{\hour} measurements comparing the unregulated and regulated magnetic field measured by the out of loop sensor. In the regulated case (blue curve), the signal of the NV control is used to actively stabilize the magnetic field. In the unregulated case (grey curve) the magnetic field is just set by the current output of the power supply. In addition to higher overall noise for the current stabilized field, jumps in the field strength created by other experiments nearby are visible here.}
\end{figure*}

In this article, we investigate the direct control of homogeneous magnetic fields based on such NV centers. Our experimental setup is sketched in Fig. \ref{fig:total}. A set of Helmholtz coils produces a homogeneous magnetic field $\Bbias$, which is directly proportional to the applied control current flowing through the coils. The magnetic field is sensed through the fluorescence of a compact NV center magnetometer. In the magnetometer, we measure the optically detected magnetic-resonance features (ODMR) by dressing the ground state spin triplet with a microwave signal of frequency $\nu$. A dip appears in the fluorescence intensity $I$ of the NV centers when $\nu$ coincides with of one the Zeeman shifted transition frequencies within the triplet \cite{OMDR_Wrachtrup}. The derivative $S = \sfrac{dI}{d\nu}$ is then used as an error signal for the feedback loop. The magnetic field strength has been monitored through a second, independent NV magnetometer over the course of 24 hours. We achieved a magnetic field stability of $\sigma_{DC} = \SI{12}{\nano \tesla}$ measured over a \SI{1}{\hertz} bandwidth, while shorter measurements over the full control loop bandwidth of \SI{1}{\kilo \hertz} yielded magnetic field noise of \SI{126}{\nano \tesla}. Replacing this second diamond with the cold atomic clouds of interest in the future will allow regulating the magnetic field the atoms experience.

The remainder of the paper is structured as follows: In section \ref{Sec:Magnetometry} we review the employed detection of the magnetic field and how we separate thermal drifts from magnetic field changes. In section \ref{sec:Stab} we discuss the control of the magnetic field and benchmark our system in a detailed fashion. We end with an outlook for more compact solutions beyond this proof-of-concept in section \ref{sec:Conc}.

\renewcommand\arraystretch{1.35}

\begin{table*}
\begin{centering}
 \begin{tabular}{| c | c | c | c | c |}
 \hline
 Sensor &  Maximum operation  field, $\si{\milli \tesla}$ & Sensitivity,  $\sfrac{\si{\nano \tesla}}{\sqrt{\si{\hertz}}}$& Blocked optical access, $\si{\milli \meter}$ & References \\
 
 \hline\hline

 Hall sensor & $ 70 - 200 $ &  $100 - 500$ & $ 1.5 -  3 $& \cite{HallSensor1, HallSensor2, HallSensor3} \\

 \hline
 
 Fluxgate  & $1$ & $ 5 \cdot 10^{-3} - 10^{-2} $ & $20 - 30$ & \cite{bookmag, Fluxgate_review, GMI} \\
 \hline
 
 Giant magnetoimpedance  & $250 \cdot 10^{-3} - 1$ & $10^{-3}-10^{-2}$ & $ 0.1 - 20 $ \footnote{Based on the orientation of the wire} & \cite{bookmag, GMI} \\
 \hline
 
 Atomic vapor cell  & $10^{-1}$ & $10^{-7}-10^{-1}$ & $2-20$ &\cite{Dang2010,Griffith2010, Mitchell_Mag} \\
 \hline
 
 SQUID  & $10^{4}$ & $10^{-6}-10^{-2}$ & $10^{-1}-20$ \footnote{Requires cryogenic cooling} & \cite{Kirtley1995,Finkler2010, bookmag, HighFieldSQUID} \\
 \hline
 \textbf{NV center} &  \boldmath{$10^4$}  & \boldmath{$9 \cdot 10^{-4} - 10$} & \boldmath{$10^{-1}-2$} & \cite{Stepanov2015, Barry2013, Walsworth_Vec, Clevenson2015, Fescenko2019}\\
 \hline
\end{tabular}
\caption{\label{Tab:Magnetometers} Comparison of commercially available Hall- and Fluxgate sensors with GMI magnetometers, atomic vapor cells, SQUIDs and NV centers. The extent of the sensor head is used to quantify the optical access to the experiment blocked.}
\end{centering}
\end{table*}

\section{Magnetometry Method}\label{Sec:Magnetometry}

\begin{figure}
\includegraphics[width = 0.45\textwidth]{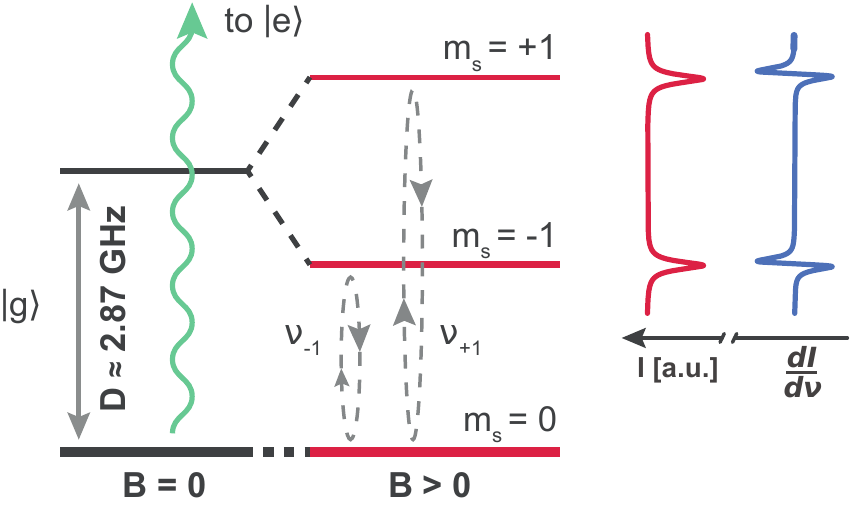}
\caption{\label{fig:fig2} Simplified NV$^{-}$ center level scheme: The internal spin state can be read out by transferring the system into the excited state - if the centers were in the $m_s = \pm 1$, the fluorescence will not be as bright (as indicated by the width of the red lines). Now the Zeeman shift of the $m_s = \pm 1$ levels can be measured by finding the fluorescence minima. By modulating the microwave frequencies $\nu_{-1}$ and $\nu_{+1}$, a lock-in scheme can be employed.}
\end{figure}

For generating the magnetic field signal on both our sensors we employ an ODMR scheme \cite{ODMR_Wrachtrup}, which we summarize in Fig.~\ref{fig:fig2}. The NV center has a spin $S=1$ in its ground state.  It is optically pumped by a green laser, and consequently emits fluorescence above \SI{637}{\nano \meter}. The unsplit $\mS = \pm 1$ states lie $D \approx \SI{2.87}{\giga \hertz}$ (also called the zero field splitting (ZFS)) above the $\mS=0$ state and hence can be coupled to the ground state through standard microwave techniques (see appendix~\ref{App:MW}). At nonzero magnetic field, the $\mS = \pm 1$ states experience a Zeeman shift of approximately $\pm h \gamma_e \vec{B}_0 \cdot \hat{n}$, where $\Bbias$ is the magnetic field strength, $\gamma_e = \SI{28}{\sfrac{\giga \hertz}{\tesla}}$ is the gyromagnetic ratio of the electron, and $\hat{n}$ is the projection of the magnetic field onto the NV center axis. For the $\mS = \pm 1$ states a nonradiative decay channel to the ground state leads to a slight decrease in fluorescence, allowing us to extract the transition frequencies $\fUp$ and $\fDown$ from the fluorescence spectrum.

We use a lock-in scheme by frequency modulating the microwave signals applied at modulation frequencies $\omega_{\text{mod$\pm$1}}$ of a few tens of \si{\kilo \hertz} to generate the error signals $\SminOne = \frac{dI}{d\nu}\rvert_{\nu = \nu_{-1}}$ and $\SplOne = \frac{dI}{d\nu}\rvert_{\nu = \nu_{+1}}$ at both transitions. The information gained by monitoring both transitions can then be used to compensate for temperature fluctuations of the sample (see appendix~\ref{App:TempControl}), which would shift both resonance frequencies in the same direction, as illustrated in Fig.~\ref{fig:total} B \cite{Temp_dep}.

The frequency modulated signal necessary for this scheme is generated on a StemLab RedPitaya, and mixed with the two channels of a microwave generator to shift them to the microwave regime to address the two transitions. After that, they are amplified and applied to a wire loop placed around the diamond sample.

The diamond samples used were grown with the HPHT technique, have a natural abundance of \textsuperscript{13}C and are type 1b diamonds with a natural linewidth of \SI{1.3(1)}{\mega \hertz}, heatsunk to a sapphire window. The samples are placed in a homogeneous magnetic field generated by a pair of Helmholtz coils, with a second pair of coils with only few windings wound onto in order to regulate the magnetic field over a high bandwidth (see appendix~\ref{App:Coils} for further details).

The fluorescence of our sample is collected efficiently by a compound parabolic concentrator \cite{CPC}, and the remaining excitation light is filtered out by a longpass filter (see appendix~\ref{App:Optics} for more details). It is measured using an auto-balanced photodetector \cite{Autobal_Patent, Autobal_Book}, whose output is also digitized by the RedPitaya. A modified version of PyRPL \cite{PyRPL_paper} is used to process the data on its FPGA, generating an error signal at both transitions to regulate the current through the low inductance coil pair.

\begin{figure}
\includegraphics[width = 0.48\textwidth]{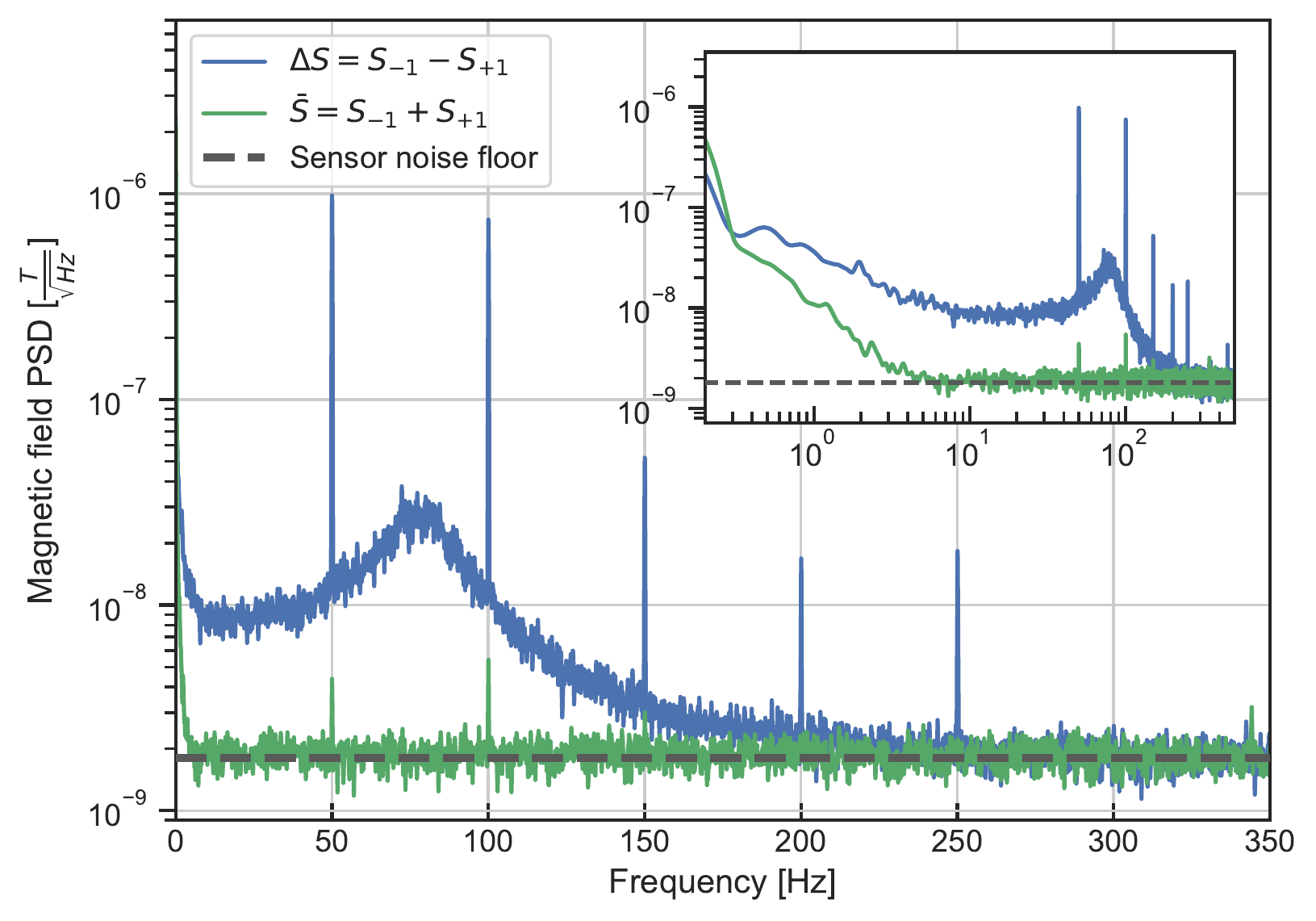}
\caption{\label{fig:temp}Sum and difference of the error signals obtained on the two transitions. The inset shows the same data plotted on a logarithmic frequency axis, indicating that the temperature noise starts to dominate at low frequencies. The dashed grey line indicates the sensor noise floor $\sigma_s$.}
\end{figure}

The position of these transitions depends on the magnetic field, but also on other factors, most importantly temperature. This effect shifts both transitions equally. As the magnetic field shift both lines in opposite directions, we only used the difference signal $\Sdiff = \SplOne- \SminOne$ as a control for the magnetic field. The sum of the two error signals $\Ssum = \SplOne + \SminOne$ is directly proportional to the shift of the ZFS $D$ and the off-axis magnetic field.

The spectrum of the signals obtained this way are shown in Fig. \ref{fig:temp}: One can see that the noise at \SI{50}{\hertz} and harmonics, which is magnetic field noise created by the power line, is by a factor 200 smaller in $\Ssum$ compared to $\Sdiff$. From this we conclude that we separate in-axis reading well from off-axis magnetic field and temperature readings, which helps us to stabilize the field in the axis of interest without introducing noise into the system.
In the inset we present the same data on a logarithmic frequency scale. Here one can clearly identify the strong temperature noise, which would eventually become dominant at low frequencies in the individual error signals $\SplOne$ and $\SminOne$.

In total, we collect about \SI{3.4}{\milli \watt} of fluorescence on the signal photodiode for both sensors, leading to a photocurrent of \SI{1.7}{\milli \ampere}. This allows us to calculate the shot noise on the photocurrent using $\sigma_{I, SN} = \sqrt{2 \cdot I \cdot \si{\elementarycharge} \cdot \Delta f}$ \cite{Walsworth_Vec}. In this equation, $I$ denotes the photocurrent, $e$ the elementary charge, and $\Delta f$ is the single-sided measurement bandwidth.

The photocurrent shot noise can be converted to the shot noise limited magnetic field sensitivity $\sigma_{B, SN} = \sigma_{I, SN} \cdot \sfrac{dB}{dI}$ by multiplying it with the inverse of the lock-in slope. This yields a shot noise limited noise floor of $\sigma_{B, SN} \approx \SI{0.97}{\sfrac{\nano \tesla}{\sqrt{\hertz}}}$ for the magnetic field signal, whereas we observe a total noise floor of $\sigma_{S} \approx \SI{1.8}{\sfrac{\nano \tesla}{\sqrt{\hertz}}}$ experimentally. The main additional contributions to this noise floor are not entirely cancelled fluorescence intensity noise as well as microwave noise, with a smaller contribution stemming from the shot noise on the reference photocurrent, which was not considered here.

\section{\label{sec:Stab}Magnetic field stabilization}

\begin{figure}
\includegraphics[width = 0.48\textwidth]{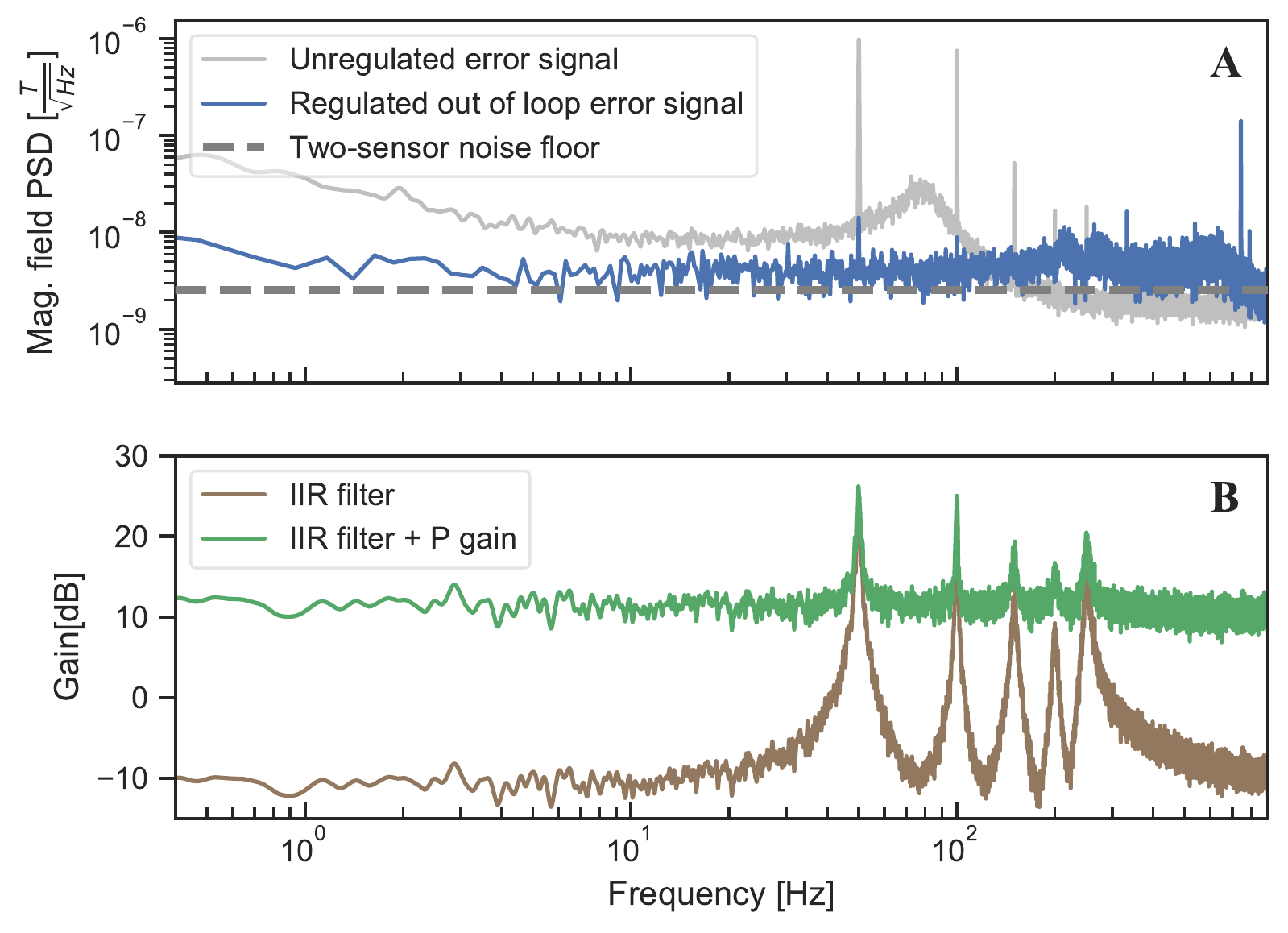}
\caption{\label{fig:lock}\textbf{(A)}: Magnetic field error signal for an open feedback loop (in grey) and for a closed feedback loop, measured on an independent sensor (in blue). \textbf{(B)}: The total feedback gain consists of a PI signal (the I part not shown) as well as a IIR filter to increase the noise suppression at \SI{50}{\hertz} and harmonics.}
\end{figure}

After the characterization of the magnetic field sensors, we employ them to monitor the magnetic field noise and actively stabilize it. The time trace of these measurements over \SI{24}{\hour} was presented in Fig.~\ref{fig:total}~C. In Fig.~\ref{fig:lock}~A we investigate the associated spectrum at high frequencies. In the open loop configuration (grey curve in the figures), the magnetic field is set to $B_0 = \SI{4.67}{\milli \tesla}$ by applying a  current of $ \approx \SI{6.3}{\ampere}$ to a pair of Helmholtz coils.
The noise spectrum in Fig.~\ref{fig:lock}~A reveals that the power supply driving this current introduces magnetic field noise of a few $  \SI{10}{\sfrac{\nano \tesla}{\sqrt{\hertz}}}$ up to a frequency of $\SI{200}{\hertz}$. Above this cutoff frequency, which is set by the inductance of the coils,  the detected magnetic field noise is again limited by the sensor noise.

We use the error signal of the magnetometer to actively stabilize the magnetic field through a feedback system and sense the resulting magnetic field on the independent sensor. The feedback is created by feeding the error signal into a PI controller (which is realized on the RedPitaya's FPGA) that regulates the current through the low inductance coil pair. Because particularly strong magnetic field noise is present at \SI{50}{\hertz} and harmonics due to power line noise, we implement an IIR filter to increase the feedback loop gain at those frequencies as shown in Fig. \ref{fig:lock}~B.

We monitor the resulting magnetic field over \SI{24}{\hour} as shown by the blue curve in Fig.~\ref{fig:total}~C. Over a bandwidth of \SI{1}{\hertz} we reduce the magnetic field noise of \SI{65}{\nano \tesla} down to $\sigma_{DC} = \SI{12}{\nano \tesla}$. The DC component of the stability is essential for the repeatability of the quantum simulation experiments, where single runs are typically performed on the timescale of up to a minute and datasets are accumulated over the duration of days \cite{LongAveraging, LongAveragingSchmiedmayer}. The spectrum of a short-term measurement covering the full stabilization bandwidth is shown in Fig.~\ref{fig:lock}~A: Here we see, that up to the control loop's bandwidth of \SI{1}{\kilo \hertz} magnetic field noise is reduced down to $\sigma_{AC} = \SI{126}{\nano \tesla}$. Within the noise spectrum, peaks that might coincide with a motional resonance are especially problematic. This makes the reduction of the \SI{50}{\Hz} noise by a factor of about $70$ a potentially crucial improvement for the fidelity of the quantum simulator.

\begin{figure}
\includegraphics[width = 0.48\textwidth]{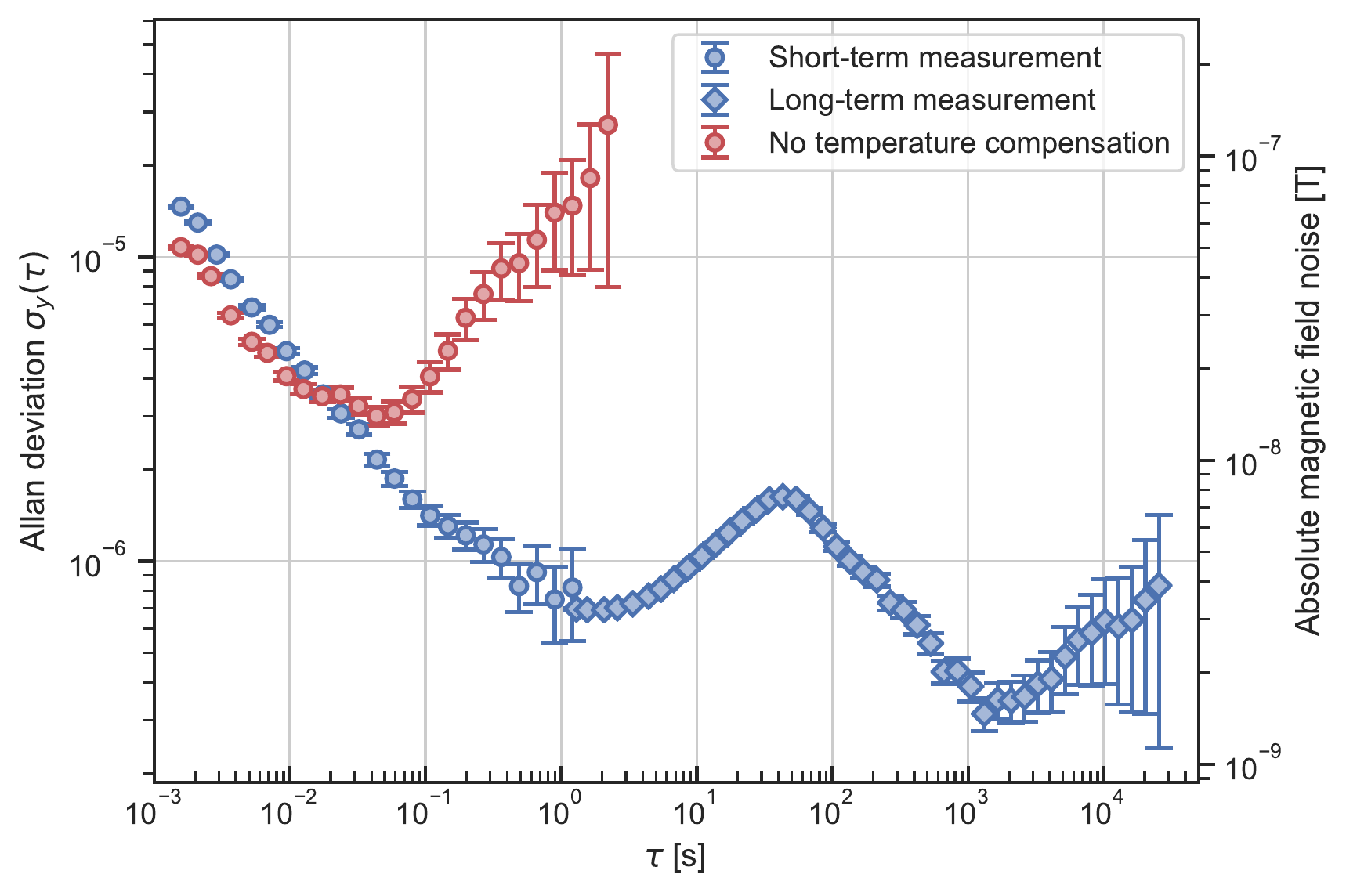}
\caption{\label{fig:adev}Allan deviation of the stabilized magnetic field. The blue diamonds are obtained from the long term measurement that is visualized in Fig.~\ref{fig:total}~C. It is complemented by a set of short term measurements (blue circles) with duration of \SI{4}{\second} and a resolution of \SI{0.2}{\milli \second} that connect directly to the long term measurements at a time scale of \SI{1}{\second}.  For up to a few seconds the Allan deviation decreases with $\sfrac{1}{\sqrt{\tau}}$ as expected for white noise on the error signal.  A noise peak at $\tau = \SI{60}{\second}$ can be observed, which results from the retuning routine for the microwave generators. The red curve shows the Allan deviation obtained when the magnetic field is only locked to $\SminOne$, so without any additional temperature noise cancellation. Its increase after approximately \SI{100}{\milli \second} is due to uncompensated temperature fluctuations, which are strongly supressed in the full scheme.}
\end{figure}

From these two measurements we calculate the Allan deviation of the magnetic field stability, as shown in Fig. \ref{fig:adev}. Additionally, a short-term measurement only stabilized to $\SminOne$, so without our temperature noise compensation scheme, is shown, clearly indicating the necessity of this scheme. The short-term measurement with temperature noise compensation drops with the $\tau^{-\sfrac{1}{2}}$ scaling expected from white frequency noise. The long-term measurements directly connects to the short-term measurement at the integration time of $\SI{1}{\second}$. For longer integration times we observe a broad peak centered at \SI{60}{\second}. This is due to the dead time during the retuning of the microwave sources once a minute to follow the temperature shifts of the NV center's transitions (see appendix~\ref{App:TempControl}). We also attribute the slight increase in magnetic field noise at low frequencies in Fig.~\ref{fig:lock}~A to these retuning effects. We hence see them as the main source of the low frequency noise $\sigma_{DC}$.

Our work extends previous work on feedback control using NV centers: In the work of \cite{Jaskula2019}, feedback from the electron spin of NV centers was used to stabilize the precession of the \textsuperscript{14}N nuclear spin ensemble associated with them by regulating the detuning of a Ramsey sequence. This allowed to reach longterm stabilities of several hours. Similarly, in \cite{Clevenson2}, a feedback loop tracks resonance pairs of NV centers by supplying feedback to the microwave frequency. This technique offered an increase in dynamic range, and made it robust against temperature fluctuations. In \cite{Glenn2018} a second NV magnetometer, which was operated close to the sample of the NV-NMR setup, was utilised to stabilize the external magnetic field. The short-term stabilization of up to 5 minutes showed a magnetic field stability of $\sigma_B \approx \SI{15}{\nano \tesla}$ over a bandwidth of $\SI{12.5}{\hertz}$. However, every 5 minutes, the in-loop sensor signal was zeroed, and the NV-NMR machine was recalibrated to avoid slow drifts of the field. This approach effectively erased information about longterm magnetic field drifts, while the microwave retuning described here only cancels temperature drifts to stay on the resonances, and thus conserves the longterm magnetic field information.

\section{\label{sec:Conc}Conclusion}

In this article we presented a direct magnetic field stabilization based on NV centers, which is compatible with intermediate magnetic fields at the typical scales in cold atom experiments. Our prototype achieved performances which are better than the state-of-the-art for quantum gas experiments at high magnetic fields. The sensitivity of the sensor can be improved further by utilizing better diamond samples, higher collection and excitation efficiencies, increasing the size of the used ensemble \cite{Clevenson2015, Barry2013, Subpico_Mag}, using magnetic flux concentrating materials \cite{Fescenko2019} or microwave cavity readout \cite{Eisenach2020,Ebel2020}. A fiber based approach to reading out the NV's spin states \cite{NVfiber, NVfiber2, Fedotov2014} could allow for an even more compact design, which would open the route to applications in confined spaces. With several magnetic field sensors positioned suitably around the cold atom chamber we could extend the scheme beyond the control of magnetic offset fields and directly control strong magnetic field gradients.

Altogether, this technology could allow for a new class of precision experiments in atomic physics at magnetic field strengths previously not accessible. One route here would be experiments on Feshbach resonances, for instance for the quantum simulation of dipolar droplets \cite{DropletsLetitia}, where the magnetic field stability would improve by up to two orders of magnitude from $\sim \SI{1}{\micro T}$ to the demonstrated $\sigma_{DC} = \SI{12}{\nano T}$, or the study of supersolid behaviors at high field values in dipolar quantum gases \cite{SupersolidFerl, SupersolidPfau}. Another possible application is the study of spin changing collisions between bosonic and fermionic species. For Bose-Bose mixtures, these resonances typically lie in the range of a hundred \si{\micro \tesla} and have been successfully employed before as a building block of abelian lattice gauge theories \cite{MilGauge, CohHeteroSpin}. However, for the simulation of quantum electrodynamics it will be necessary to employ fermionic atoms, which emulate the role of the fermionic charges. For the scattering of a bosonic and a fermionic species they occur at much higher field strengths, e.g. for the case of $^{23}\mathrm{Na}$ and $^{40}\mathrm{K}$ in the vicinity of \SI{30}{\milli \tesla}. The field stability $\sigma_{DC}$ results in an energy stability in the order of $\SI{100}{\hertz}$ here, which is putting the quantum simulation of quantum electrodynamics into an accessible regime \cite{Zache2018}.

Finally, the hybrid of ultracold quantum gases and color centers might allow for a new class of experiments at the interface of these two quantum systems. One possible application here would be the cross validation of the previously measured g-factor of the NV center \cite{Loubser1978, Doherty2011,Zhukov2017} with higher precision than done before.

\begin{acknowledgments}
This work is part of and supported by the DFG Collaborative Research Centre ``SFB 1225 (ISOQUANT)''. F. J. acknowledges  the DFG support through the project FOR 2724, the Emmy-Noether grant (project-id 377616843) and from the Juniorprofessorenprogramm Baden-W\"urttemberg (MWK). D.D. and J.W.  acknowledge financial support by the  German Science Foundation (DFG) (SPP1601, FOR2724),  the EU (ERC) (ASTERIQS, SMel),  the Max Planck Society, the Volkswagen Stiftung. We thank Helmut Strobel for helpful discussions and Leonhard Neuhaus for advice on modifying PyRPL's FPGA.
The raw data and the data treatment are accessible online \cite{Heidata}.

\end{acknowledgments}

\appendix

\begin{figure*}[t]
\includegraphics[width = \textwidth]{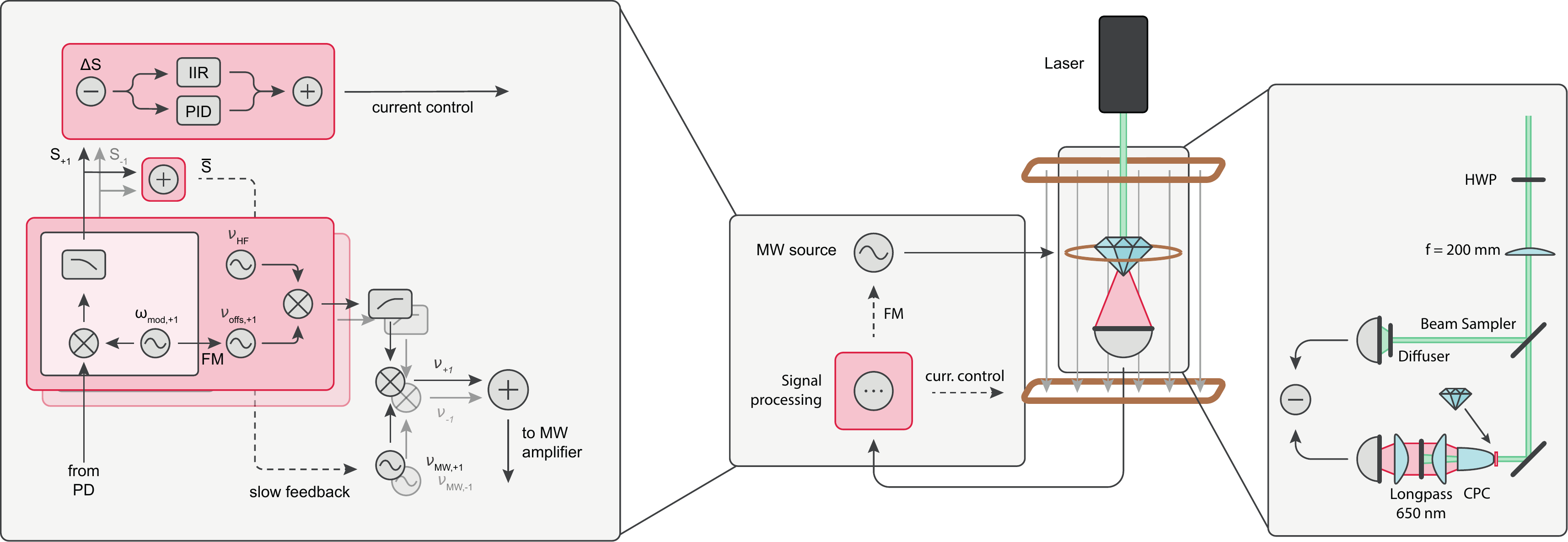}
\caption{\label{fig:setup}Overviev over the experimental setup, with a breakout for the microwave setup and control loop on the left, and the optical setup on the right. All elements enclosed in a red box are realized on the FPGA of a StemLab RedPitaya, with the lock in amplifier components grouped in a lighter colored box.}
\end{figure*}

\section{Optical setup}\label{App:Optics}

For both sensors used in this work the optical setups are identical, with exception of the laser: Either a Laser Quantum Finesse laser or a \SI{520}{\nano \meter} diode laser acquired off ebay supply approximately \SI{900}{\milli \watt} of excitation light. We focus it down onto the diamond samples using a f=\SI{200}{\milli \meter} lens, and collect the fluorescence using a compound parabolic concentrator (Edmund Optics, with a 25\degree collection angle and a \SI{4.3}{\milli \meter} exit diameter). The fluorescence passes a \SI{650}{\nano \meter} longpass filter (Thorlabs FEL0650) to filter out any remaining excitation light before being focused onto the signal photodiode of an autobalanced photodetector (Nirvana Auto-Balanced Photoreceivers Model 2007).

To collect a reference beam for our photodetector, the excitation beam is sent through a $\sfrac{\lambda}{2}$ waveplate as well as a beam sampler (Thorlabs BSF10-A). The sampled beam passes through a diffuser (Thorlabs DG10-220-A) to remove etalon fringing and similar effects before being detected and used as the reference on the autobalanced photodetector.

\section{Microwave setup and control loop}\label{App:MW}

In order to keep both the microwave setup and the control loop (illustrated on the left in Fig \ref{fig:setup}) compact, much of the functionality was realized on the FPGA of a Red Pitaya STEMlab 125-14 board using a modified version of PyRPL \cite{PyRPL_paper}. This allows us to use only few additional components to shift the frequencies into the \si{\giga \hertz} regime.

To cancel out the temperature noise in our signal, both transitions $\mS = 0 \rightarrow \mS = -1$ and $\mS = 0 \rightarrow \mS = +1$ are monitored simultaneously. This is possible by frequency modulating the microwave signals driving the transitions at different frequencies $\omega_{\text{mod,-1}}$ and $\omega_{\text{mod,+1}}$. The two signals are separated by demodulating the fluorescence at said frequencies using a lock-in scheme.

The frequency modulation is generated on a frequency $\nu_{\text{offs,-1}}$ as well as $\nu_{\text{offs,+1}}$ directly on the RedPitaya's FPGA. Afterwards, the signals are mixed with the NV center's hyperfine splitting $\nu_{\text{HF}}$ to generate sidebands at $\approx \SI{2.2}{\mega \hertz}$. These signals are then outputted via the RedPitaya's fast DACs, and high pass filtered afterwards (using Mini-Circuits ZFHP-0R50-S+ high pass filters) to remove any low frequency output (due to technical constraints, the current control signal is outputted on the same DAC). This filtered signal is then mixed (Mini-Circuits ZX05-43H-S+) with the microwave frequencies $\nu_{\text{MW, -1}}$ and $\nu_{\text{MW, +1}}$ (generated by the two channels of a Windfreaktech SynthHD) to shift the frequencies to the \si{\giga \hertz} regime, creating the frequencies $\nu_{\text{-1}}$ and $\nu_{\text{+1}}$ . Finally, the two signals are added together  using a power splitter (Mini-Circuits ZX10-2-442-S+) and fed into an amplifier (a Mini Circuits ZHL-16W-43-S+ in-loop or a Mini-Circuits ZHL-15W-422+ out of loop). The amplifier output is applied to the diamond samples via a wire loop glued to the same sapphire glass heatsink as the sample.	
		
Meanwhile, the output of the two lock-in modules $\SminOne$ and $\SplOne$ on the RedPitaya is normalized and subtracted (added) together to generate the signals $\Sdiff$ ($\Ssum$). $\Ssum$ is used in the temperature noise cancellation scheme explained in the next section, while $\Sdiff$ is fed into both a PI controller as well as into an IIR filter programmed to amplify noise at \SI{50}{\hertz} and harmonics to achieve a stronger noise suppression at those frequencies. Both outputs are added together and outputted on one of the RedPitaya's fast DACs. This control signal is then fed into a home built current controller regulating the current through a low inductance coil pair, and thus stabilizing the magnetic field.

\section{Temperature noise cancellation scheme}\label{App:TempControl}

Great care has to be taken in order to extract an error signal only containing the magnetic field reading along the axis of interest, and not any off-axis components or temperature fluctuation signals. Conveniently, both off-axis fluctuations as well as temperature fluctuations move the two transitions $\mS = 0 \rightarrow \mS = -1$ and $\mS = 0 \rightarrow \mS = 1$ in unison, while the on axis fluctuations move them in opposite directions, as illustrated in Fig. \ref{fig:total} B. This allows to extract the on-axis magnetic field reading by subtracting the normalized error signals generated on both transitions independently. To normalize these signals, magnetic field fluctuations at a frequency with a low noise floor were generated on the feedback coil pair, and their effect on the sum signal $\Ssum$ was minimized by scaling the two error signals accordingly.	
		
One issue that can occur in this configuration is that the temperature drifts far enough to shift the transitions away from the two microwave frequencies $\nu_{-1}$ and $\nu_{+1}$ - luckily, this can be circumvented by observing the sum signal $\Ssum$ and retuning the absolute microwave frequencies (while keeping their relative distance, which sets the magnetic field strength), keeping $\Ssum$ near zero.
	
A crude feedback loop - consisting of several measurements of $\Ssum$ followed by a retuning of the microwave frequencies in order to regulate the sum signal - was implemented for the measurements presented in this work. Even though it generates some noise for every retuning (as can be seen on the peak at \SI{60}{\second} in Fig. \ref{fig:adev}), it is essential to ensure the long-term functionality of the feedback loop, and to prevent it from falling out of lock.

\section{Magnetic field generation}\label{App:Coils}

We employ a pair of coils of side lengths $\SI{33.5}{\centi \meter} \times \SI{28.5}{\centi \meter}$ with 80 windings each and separated by \SI{15.5}{\centi \meter} to generate our magnetic bias field. A second coil pair with only 16 windings wrapped around the first coil pair is used for regulating the magnetic field over a larger bandwidth. While the bias coil pair is driven by a Delta Elektronika SM 18-50, this coil pair is regulated using a homebuilt current driver.

\begin{figure}
\includegraphics[width = 0.48\textwidth]{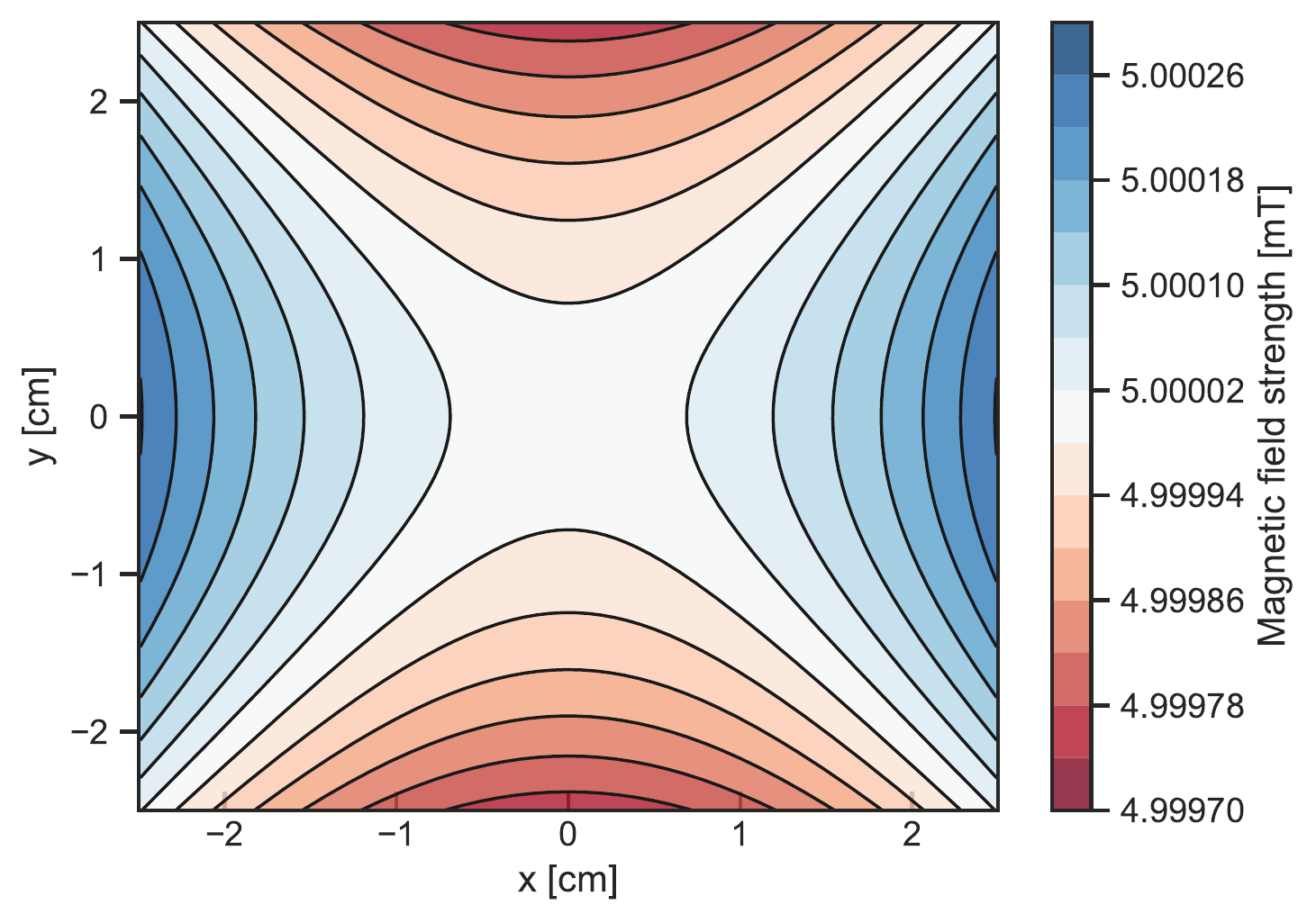}
\caption{\label{fig:B_sim} Magnetic field homogeneity of the Helmholtz coil pair in the xy-plane with z = 0. The y-axis is oriented along the Helmholtz coil axis, and spans the horizontal plane together with the x-axis.}
\end{figure}

One requirement for the magnetic field in our application is to be homogeneous over both sensors to ensure that the signals read out are identical. For this, the coil geometry used was simulated using Radia \cite{Radia}, yielding inhomogeneities below \SI{50}{ppm} for \SI{1}{\centi \meter} total misalignment of the two sensors with respect to the coil pair's center. As the AC magnetic field fluctuations we want to stabilize are on the order of \SI{1}{\sfrac{\micro \tesla}{\sqrt{\hertz}}} and below, this is currently not limiting. However, magnetic field gradients leading to different sensor readings can also be generated by external sources, like \SI{50}{\hertz} power line noise, making it desirable to keep both sensors as close together as possible.

In a similar manner, the magnetic field inside the diamond samples was simulated. As inhomogenities here lead to a broadening of the resonances for large magnetic fields, the sensitivity $S$ can be calculated using the magnetic field spread $\Delta B$ and the system's transition linewidth at zero magnetic field $\Delta f$:

\begin{equation}
S \propto \sqrt{\frac{\Delta f}{\sqrt{\Delta f^2 + \Delta B^2 \cdot \gamma_e^2}}}
\end{equation}

Based on this scaling, we expect the sensitivity of our sensor to decrease by 50 \% past a field value of \SI{2.2}{\tesla}.

\bibliography{bibliography}

\providecommand{\noopsort}[1]{}\providecommand{\singleletter}[1]{#1}%
\begin{thebibliography}{52}%
\makeatletter
\providecommand \@ifxundefined [1]{%
 \@ifx{#1\undefined}
}%
\providecommand \@ifnum [1]{%
 \ifnum #1\expandafter \@firstoftwo
 \else \expandafter \@secondoftwo
 \fi
}%
\providecommand \@ifx [1]{%
 \ifx #1\expandafter \@firstoftwo
 \else \expandafter \@secondoftwo
 \fi
}%
\providecommand \natexlab [1]{#1}%
\providecommand \enquote  [1]{``#1''}%
\providecommand \bibnamefont  [1]{#1}%
\providecommand \bibfnamefont [1]{#1}%
\providecommand \citenamefont [1]{#1}%
\providecommand \href@noop [0]{\@secondoftwo}%
\providecommand \href [0]{\begingroup \@sanitize@url \@href}%
\providecommand \@href[1]{\@@startlink{#1}\@@href}%
\providecommand \@@href[1]{\endgroup#1\@@endlink}%
\providecommand \@sanitize@url [0]{\catcode `\\12\catcode `\$12\catcode
  `\&12\catcode `\#12\catcode `\^12\catcode `\_12\catcode `\%12\relax}%
\providecommand \@@startlink[1]{}%
\providecommand \@@endlink[0]{}%
\providecommand \url  [0]{\begingroup\@sanitize@url \@url }%
\providecommand \@url [1]{\endgroup\@href {#1}{\urlprefix }}%
\providecommand \urlprefix  [0]{URL }%
\providecommand \Eprint [0]{\href }%
\providecommand \doibase [0]{http://dx.doi.org/}%
\providecommand \selectlanguage [0]{\@gobble}%
\providecommand \bibinfo  [0]{\@secondoftwo}%
\providecommand \bibfield  [0]{\@secondoftwo}%
\providecommand \translation [1]{[#1]}%
\providecommand \BibitemOpen [0]{}%
\providecommand \bibitemStop [0]{}%
\providecommand \bibitemNoStop [0]{.\EOS\space}%
\providecommand \EOS [0]{\spacefactor3000\relax}%
\providecommand \BibitemShut  [1]{\csname bibitem#1\endcsname}%
\let\auto@bib@innerbib\@empty
\bibitem [{\citenamefont {Strobel}\ \emph {et~al.}(2014)\citenamefont
  {Strobel}, \citenamefont {Muessel}, \citenamefont {Linnemann}, \citenamefont
  {Zibold}, \citenamefont {Hume}, \citenamefont {Pezz{\`e}}, \citenamefont
  {Smerzi},\ and\ \citenamefont {Oberthaler}}]{StrobelFisher}%
  \BibitemOpen
  \bibfield  {author} {\bibinfo {author} {\bibfnamefont {H.}~\bibnamefont
  {Strobel}}, \bibinfo {author} {\bibfnamefont {W.}~\bibnamefont {Muessel}},
  \bibinfo {author} {\bibfnamefont {D.}~\bibnamefont {Linnemann}}, \bibinfo
  {author} {\bibfnamefont {T.}~\bibnamefont {Zibold}}, \bibinfo {author}
  {\bibfnamefont {D.~B.}\ \bibnamefont {Hume}}, \bibinfo {author}
  {\bibfnamefont {L.}~\bibnamefont {Pezz{\`e}}}, \bibinfo {author}
  {\bibfnamefont {A.}~\bibnamefont {Smerzi}}, \ and\ \bibinfo {author}
  {\bibfnamefont {M.~K.}\ \bibnamefont {Oberthaler}},\ }\href {\doibase
  10.1126/science.1250147} {\bibfield  {journal} {\bibinfo  {journal}
  {Science}\ }\textbf {\bibinfo {volume} {345}},\ \bibinfo {pages} {424}
  (\bibinfo {year} {2014})}\BibitemShut {NoStop}%
\bibitem [{\citenamefont {Muessel}\ \emph
  {et~al.}(2014{\natexlab{a}})\citenamefont {Muessel}, \citenamefont {Strobel},
  \citenamefont {Linnemann}, \citenamefont {Hume},\ and\ \citenamefont
  {Oberthaler}}]{MuesselSqueezing}%
  \BibitemOpen
  \bibfield  {author} {\bibinfo {author} {\bibfnamefont {W.}~\bibnamefont
  {Muessel}}, \bibinfo {author} {\bibfnamefont {H.}~\bibnamefont {Strobel}},
  \bibinfo {author} {\bibfnamefont {D.}~\bibnamefont {Linnemann}}, \bibinfo
  {author} {\bibfnamefont {D.~B.}\ \bibnamefont {Hume}}, \ and\ \bibinfo
  {author} {\bibfnamefont {M.~K.}\ \bibnamefont {Oberthaler}},\ }\href
  {\doibase 10.1103/PhysRevLett.113.103004} {\bibfield  {journal} {\bibinfo
  {journal} {Phys. Rev. Lett.}\ }\textbf {\bibinfo {volume} {113}},\ \bibinfo
  {pages} {103004} (\bibinfo {year} {2014}{\natexlab{a}})}\BibitemShut
  {NoStop}%
\bibitem [{\citenamefont {Pr{\"u}fer}\ \emph {et~al.}(2018)\citenamefont
  {Pr{\"u}fer}, \citenamefont {Kunkel}, \citenamefont {Strobel}, \citenamefont
  {Lannig}, \citenamefont {Linnemann}, \citenamefont {Schmied}, \citenamefont
  {Berges}, \citenamefont {Gasenzer},\ and\ \citenamefont
  {Oberthaler}}]{MaxNature}%
  \BibitemOpen
  \bibfield  {author} {\bibinfo {author} {\bibfnamefont {M.}~\bibnamefont
  {Pr{\"u}fer}}, \bibinfo {author} {\bibfnamefont {P.}~\bibnamefont {Kunkel}},
  \bibinfo {author} {\bibfnamefont {H.}~\bibnamefont {Strobel}}, \bibinfo
  {author} {\bibfnamefont {S.}~\bibnamefont {Lannig}}, \bibinfo {author}
  {\bibfnamefont {D.}~\bibnamefont {Linnemann}}, \bibinfo {author}
  {\bibfnamefont {C.-M.}\ \bibnamefont {Schmied}}, \bibinfo {author}
  {\bibfnamefont {J.}~\bibnamefont {Berges}}, \bibinfo {author} {\bibfnamefont
  {T.}~\bibnamefont {Gasenzer}}, \ and\ \bibinfo {author} {\bibfnamefont
  {M.~K.}\ \bibnamefont {Oberthaler}},\ }\href {\doibase
  10.1038/s41586-018-0659-0} {\bibfield  {journal} {\bibinfo  {journal}
  {Nature}\ }\textbf {\bibinfo {volume} {563}},\ \bibinfo {pages} {217}
  (\bibinfo {year} {2018})}\BibitemShut {NoStop}%
\bibitem [{\citenamefont {Mil}\ \emph {et~al.}(2020)\citenamefont {Mil},
  \citenamefont {Zache}, \citenamefont {Hegde}, \citenamefont {Xia},
  \citenamefont {Bhatt}, \citenamefont {Oberthaler}, \citenamefont {Hauke},
  \citenamefont {Berges},\ and\ \citenamefont {Jendrzejewski}}]{MilGauge}%
  \BibitemOpen
  \bibfield  {author} {\bibinfo {author} {\bibfnamefont {A.}~\bibnamefont
  {Mil}}, \bibinfo {author} {\bibfnamefont {T.~V.}\ \bibnamefont {Zache}},
  \bibinfo {author} {\bibfnamefont {A.}~\bibnamefont {Hegde}}, \bibinfo
  {author} {\bibfnamefont {A.}~\bibnamefont {Xia}}, \bibinfo {author}
  {\bibfnamefont {R.~P.}\ \bibnamefont {Bhatt}}, \bibinfo {author}
  {\bibfnamefont {M.~K.}\ \bibnamefont {Oberthaler}}, \bibinfo {author}
  {\bibfnamefont {P.}~\bibnamefont {Hauke}}, \bibinfo {author} {\bibfnamefont
  {J.}~\bibnamefont {Berges}}, \ and\ \bibinfo {author} {\bibfnamefont
  {F.}~\bibnamefont {Jendrzejewski}},\ }\href {\doibase
  10.1126/science.aaz5312} {\bibfield  {journal} {\bibinfo  {journal}
  {Science}\ }\textbf {\bibinfo {volume} {367}},\ \bibinfo {pages} {1128}
  (\bibinfo {year} {2020})}\BibitemShut {NoStop}%
\bibitem [{\citenamefont {Ripka}(1992)}]{Fluxgate_review}%
  \BibitemOpen
  \bibfield  {author} {\bibinfo {author} {\bibfnamefont {P.}~\bibnamefont
  {Ripka}},\ }\href {\doibase https://doi.org/10.1016/0924-4247(92)80159-Z}
  {\bibfield  {journal} {\bibinfo  {journal} {Sensors and Actuators A:
  Physical}\ }\textbf {\bibinfo {volume} {33}},\ \bibinfo {pages} {129 }
  (\bibinfo {year} {1992})}\BibitemShut {NoStop}%
\bibitem [{\citenamefont {Cabrera}\ \emph {et~al.}(2018)\citenamefont
  {Cabrera}, \citenamefont {Tanzi}, \citenamefont {Sanz}, \citenamefont
  {Naylor}, \citenamefont {Thomas}, \citenamefont {Cheiney},\ and\
  \citenamefont {Tarruell}}]{DropletsLetitia}%
  \BibitemOpen
  \bibfield  {author} {\bibinfo {author} {\bibfnamefont {C.~R.}\ \bibnamefont
  {Cabrera}}, \bibinfo {author} {\bibfnamefont {L.}~\bibnamefont {Tanzi}},
  \bibinfo {author} {\bibfnamefont {J.}~\bibnamefont {Sanz}}, \bibinfo {author}
  {\bibfnamefont {B.}~\bibnamefont {Naylor}}, \bibinfo {author} {\bibfnamefont
  {P.}~\bibnamefont {Thomas}}, \bibinfo {author} {\bibfnamefont
  {P.}~\bibnamefont {Cheiney}}, \ and\ \bibinfo {author} {\bibfnamefont
  {L.}~\bibnamefont {Tarruell}},\ }\href {\doibase 10.1126/science.aao5686}
  {\bibfield  {journal} {\bibinfo  {journal} {Science}\ }\textbf {\bibinfo
  {volume} {359}},\ \bibinfo {pages} {301} (\bibinfo {year}
  {2018})}\BibitemShut {NoStop}%
\bibitem [{\citenamefont {Zache}\ \emph {et~al.}(2018)\citenamefont {Zache},
  \citenamefont {Hebenstreit}, \citenamefont {Jendrzejewski}, \citenamefont
  {Oberthaler}, \citenamefont {Berges},\ and\ \citenamefont
  {Hauke}}]{Zache2018}%
  \BibitemOpen
  \bibfield  {author} {\bibinfo {author} {\bibfnamefont {T.~V.}\ \bibnamefont
  {Zache}}, \bibinfo {author} {\bibfnamefont {F.}~\bibnamefont {Hebenstreit}},
  \bibinfo {author} {\bibfnamefont {F.}~\bibnamefont {Jendrzejewski}}, \bibinfo
  {author} {\bibfnamefont {M.~K.}\ \bibnamefont {Oberthaler}}, \bibinfo
  {author} {\bibfnamefont {J.}~\bibnamefont {Berges}}, \ and\ \bibinfo {author}
  {\bibfnamefont {P.}~\bibnamefont {Hauke}},\ }\href {\doibase
  10.1088/2058-9565/aac33b} {\bibfield  {journal} {\bibinfo  {journal} {Quantum
  Science and Technology}\ }\textbf {\bibinfo {volume} {3}},\ \bibinfo {pages}
  {034010} (\bibinfo {year} {2018})}\BibitemShut {NoStop}%
\bibitem [{\citenamefont {Baudenbacher}\ \emph {et~al.}(2003)\citenamefont
  {Baudenbacher}, \citenamefont {Fong de~los Santos}, \citenamefont {Holzer},\
  and\ \citenamefont {Radparvar}}]{SQUID1}%
  \BibitemOpen
  \bibfield  {author} {\bibinfo {author} {\bibfnamefont {F.}~\bibnamefont
  {Baudenbacher}}, \bibinfo {author} {\bibfnamefont {L.}~\bibnamefont {Fong
  de~los Santos}}, \bibinfo {author} {\bibfnamefont {J.}~\bibnamefont
  {Holzer}}, \ and\ \bibinfo {author} {\bibfnamefont {M.}~\bibnamefont
  {Radparvar}},\ }\href {\doibase 10.1063/1.1572968} {\bibfield  {journal}
  {\bibinfo  {journal} {Appl. Phys. Lett.}\ }\textbf {\bibinfo {volume} {82}},\
  \bibinfo {pages} {3487} (\bibinfo {year} {2003})}\BibitemShut {NoStop}%
\bibitem [{\citenamefont {{Drung}}\ \emph {et~al.}(2007)\citenamefont
  {{Drung}}, \citenamefont {{Abmann}}, \citenamefont {{Beyer}}, \citenamefont
  {{Kirste}}, \citenamefont {{Peters}}, \citenamefont {{Ruede}},\ and\
  \citenamefont {{Schurig}}}]{SQUID2}%
  \BibitemOpen
  \bibfield  {author} {\bibinfo {author} {\bibfnamefont {D.}~\bibnamefont
  {{Drung}}}, \bibinfo {author} {\bibfnamefont {C.}~\bibnamefont {{Abmann}}},
  \bibinfo {author} {\bibfnamefont {J.}~\bibnamefont {{Beyer}}}, \bibinfo
  {author} {\bibfnamefont {A.}~\bibnamefont {{Kirste}}}, \bibinfo {author}
  {\bibfnamefont {M.}~\bibnamefont {{Peters}}}, \bibinfo {author}
  {\bibfnamefont {F.}~\bibnamefont {{Ruede}}}, \ and\ \bibinfo {author}
  {\bibfnamefont {T.}~\bibnamefont {{Schurig}}},\ }\href {\doibase
  10.1109/TASC.2007.897403} {\bibfield  {journal} {\bibinfo  {journal} {IEEE
  Transactions on Applied Superconductivity}\ }\textbf {\bibinfo {volume}
  {17}},\ \bibinfo {pages} {699} (\bibinfo {year} {2007})}\BibitemShut
  {NoStop}%
\bibitem [{\citenamefont {Dang}\ \emph {et~al.}(2010)\citenamefont {Dang},
  \citenamefont {Maloof},\ and\ \citenamefont {Romalis}}]{Dang2010}%
  \BibitemOpen
  \bibfield  {author} {\bibinfo {author} {\bibfnamefont {H.~B.}\ \bibnamefont
  {Dang}}, \bibinfo {author} {\bibfnamefont {A.~C.}\ \bibnamefont {Maloof}}, \
  and\ \bibinfo {author} {\bibfnamefont {M.~V.}\ \bibnamefont {Romalis}},\
  }\href {\doibase 10.1063/1.3491215} {\bibfield  {journal} {\bibinfo
  {journal} {Applied Physics Letters}\ }\textbf {\bibinfo {volume} {97}},\
  \bibinfo {pages} {151110} (\bibinfo {year} {2010})}\BibitemShut {NoStop}%
\bibitem [{\citenamefont {Griffith}\ \emph {et~al.}(2010)\citenamefont
  {Griffith}, \citenamefont {Knappe},\ and\ \citenamefont
  {Kitching}}]{Griffith2010}%
  \BibitemOpen
  \bibfield  {author} {\bibinfo {author} {\bibfnamefont {W.~C.}\ \bibnamefont
  {Griffith}}, \bibinfo {author} {\bibfnamefont {S.}~\bibnamefont {Knappe}}, \
  and\ \bibinfo {author} {\bibfnamefont {J.}~\bibnamefont {Kitching}},\ }\href
  {\doibase 10.1364/OE.18.027167} {\bibfield  {journal} {\bibinfo  {journal}
  {Optics Express}\ }\textbf {\bibinfo {volume} {18}},\ \bibinfo {pages}
  {27167} (\bibinfo {year} {2010})}\BibitemShut {NoStop}%
\bibitem [{\citenamefont {Lucivero}\ \emph {et~al.}(2014)\citenamefont
  {Lucivero}, \citenamefont {Anielski}, \citenamefont {Gawlik},\ and\
  \citenamefont {Mitchell}}]{Mitchell_Mag}%
  \BibitemOpen
  \bibfield  {author} {\bibinfo {author} {\bibfnamefont {V.~G.}\ \bibnamefont
  {Lucivero}}, \bibinfo {author} {\bibfnamefont {P.}~\bibnamefont {Anielski}},
  \bibinfo {author} {\bibfnamefont {W.}~\bibnamefont {Gawlik}}, \ and\ \bibinfo
  {author} {\bibfnamefont {M.~W.}\ \bibnamefont {Mitchell}},\ }\href {\doibase
  10.1063/1.4901588} {\bibfield  {journal} {\bibinfo  {journal} {Review of
  Scientific Instruments}\ }\textbf {\bibinfo {volume} {85}},\ \bibinfo {pages}
  {113108} (\bibinfo {year} {2014})}\BibitemShut {NoStop}%
\bibitem [{\citenamefont {Muessel}\ \emph
  {et~al.}(2014{\natexlab{b}})\citenamefont {Muessel}, \citenamefont {Strobel},
  \citenamefont {Linnemann}, \citenamefont {Hume},\ and\ \citenamefont
  {Oberthaler}}]{Magnetom_Oberth}%
  \BibitemOpen
  \bibfield  {author} {\bibinfo {author} {\bibfnamefont {W.}~\bibnamefont
  {Muessel}}, \bibinfo {author} {\bibfnamefont {H.}~\bibnamefont {Strobel}},
  \bibinfo {author} {\bibfnamefont {D.}~\bibnamefont {Linnemann}}, \bibinfo
  {author} {\bibfnamefont {D.~B.}\ \bibnamefont {Hume}}, \ and\ \bibinfo
  {author} {\bibfnamefont {M.~K.}\ \bibnamefont {Oberthaler}},\ }\href
  {\doibase 10.1103/PhysRevLett.113.103004} {\bibfield  {journal} {\bibinfo
  {journal} {Phys. Rev. Lett.}\ }\textbf {\bibinfo {volume} {113}},\ \bibinfo
  {pages} {103004} (\bibinfo {year} {2014}{\natexlab{b}})}\BibitemShut
  {NoStop}%
\bibitem [{\citenamefont {Wolf}\ \emph {et~al.}(2015)\citenamefont {Wolf},
  \citenamefont {Neumann}, \citenamefont {Nakamura}, \citenamefont {Sumiya},
  \citenamefont {Ohshima}, \citenamefont {Isoya},\ and\ \citenamefont
  {Wrachtrup}}]{Subpico_Mag}%
  \BibitemOpen
  \bibfield  {author} {\bibinfo {author} {\bibfnamefont {T.}~\bibnamefont
  {Wolf}}, \bibinfo {author} {\bibfnamefont {P.}~\bibnamefont {Neumann}},
  \bibinfo {author} {\bibfnamefont {K.}~\bibnamefont {Nakamura}}, \bibinfo
  {author} {\bibfnamefont {H.}~\bibnamefont {Sumiya}}, \bibinfo {author}
  {\bibfnamefont {T.}~\bibnamefont {Ohshima}}, \bibinfo {author} {\bibfnamefont
  {J.}~\bibnamefont {Isoya}}, \ and\ \bibinfo {author} {\bibfnamefont
  {J.}~\bibnamefont {Wrachtrup}},\ }\href {\doibase 10.1103/PhysRevX.5.041001}
  {\bibfield  {journal} {\bibinfo  {journal} {Phys. Rev. X}\ }\textbf {\bibinfo
  {volume} {5}},\ \bibinfo {pages} {041001} (\bibinfo {year}
  {2015})}\BibitemShut {NoStop}%
\bibitem [{\citenamefont {Barry}\ \emph {et~al.}(2016)\citenamefont {Barry},
  \citenamefont {Turner}, \citenamefont {Schloss}, \citenamefont {Glenn},
  \citenamefont {Song}, \citenamefont {Lukin}, \citenamefont {Park},\ and\
  \citenamefont {Walsworth}}]{Barry2013}%
  \BibitemOpen
  \bibfield  {author} {\bibinfo {author} {\bibfnamefont {J.~F.}\ \bibnamefont
  {Barry}}, \bibinfo {author} {\bibfnamefont {M.~J.}\ \bibnamefont {Turner}},
  \bibinfo {author} {\bibfnamefont {J.~M.}\ \bibnamefont {Schloss}}, \bibinfo
  {author} {\bibfnamefont {D.~R.}\ \bibnamefont {Glenn}}, \bibinfo {author}
  {\bibfnamefont {Y.}~\bibnamefont {Song}}, \bibinfo {author} {\bibfnamefont
  {M.~D.}\ \bibnamefont {Lukin}}, \bibinfo {author} {\bibfnamefont
  {H.}~\bibnamefont {Park}}, \ and\ \bibinfo {author} {\bibfnamefont {R.~L.}\
  \bibnamefont {Walsworth}},\ }\href {\doibase 10.1073/pnas.1601513113}
  {\bibfield  {journal} {\bibinfo  {journal} {Proceedings of the National
  Academy of Sciences}\ }\textbf {\bibinfo {volume} {113}},\ \bibinfo {pages}
  {14133} (\bibinfo {year} {2016})}\BibitemShut {NoStop}%
\bibitem [{\citenamefont {Schloss}\ \emph {et~al.}(2018)\citenamefont
  {Schloss}, \citenamefont {Barry}, \citenamefont {Turner},\ and\ \citenamefont
  {Walsworth}}]{Walsworth_Vec}%
  \BibitemOpen
  \bibfield  {author} {\bibinfo {author} {\bibfnamefont {J.~M.}\ \bibnamefont
  {Schloss}}, \bibinfo {author} {\bibfnamefont {J.~F.}\ \bibnamefont {Barry}},
  \bibinfo {author} {\bibfnamefont {M.~J.}\ \bibnamefont {Turner}}, \ and\
  \bibinfo {author} {\bibfnamefont {R.~L.}\ \bibnamefont {Walsworth}},\ }\href
  {\doibase 10.1103/PhysRevApplied.10.034044} {\bibfield  {journal} {\bibinfo
  {journal} {Phys. Rev. Applied}\ }\textbf {\bibinfo {volume} {10}},\ \bibinfo
  {pages} {034044} (\bibinfo {year} {2018})}\BibitemShut {NoStop}%
\bibitem [{\citenamefont {Clevenson}\ \emph {et~al.}(2015)\citenamefont
  {Clevenson}, \citenamefont {Trusheim}, \citenamefont {Teale}, \citenamefont
  {Schr{\"{o}}der}, \citenamefont {Braje},\ and\ \citenamefont
  {Englund}}]{Clevenson2015}%
  \BibitemOpen
  \bibfield  {author} {\bibinfo {author} {\bibfnamefont {H.}~\bibnamefont
  {Clevenson}}, \bibinfo {author} {\bibfnamefont {M.~E.}\ \bibnamefont
  {Trusheim}}, \bibinfo {author} {\bibfnamefont {C.}~\bibnamefont {Teale}},
  \bibinfo {author} {\bibfnamefont {T.}~\bibnamefont {Schr{\"{o}}der}},
  \bibinfo {author} {\bibfnamefont {D.}~\bibnamefont {Braje}}, \ and\ \bibinfo
  {author} {\bibfnamefont {D.}~\bibnamefont {Englund}},\ }\href {\doibase
  10.1038/nphys3291} {\bibfield  {journal} {\bibinfo  {journal} {Nature
  Physics}\ }\textbf {\bibinfo {volume} {11}},\ \bibinfo {pages} {393}
  (\bibinfo {year} {2015})}\BibitemShut {NoStop}%
\bibitem [{\citenamefont {Fescenko}\ \emph {et~al.}(2020)\citenamefont
  {Fescenko}, \citenamefont {Jarmola}, \citenamefont {Savukov}, \citenamefont
  {Kehayias}, \citenamefont {Smits}, \citenamefont {Damron}, \citenamefont
  {Ristoff}, \citenamefont {Mosavian},\ and\ \citenamefont
  {Acosta}}]{Fescenko2019}%
  \BibitemOpen
  \bibfield  {author} {\bibinfo {author} {\bibfnamefont {I.}~\bibnamefont
  {Fescenko}}, \bibinfo {author} {\bibfnamefont {A.}~\bibnamefont {Jarmola}},
  \bibinfo {author} {\bibfnamefont {I.}~\bibnamefont {Savukov}}, \bibinfo
  {author} {\bibfnamefont {P.}~\bibnamefont {Kehayias}}, \bibinfo {author}
  {\bibfnamefont {J.}~\bibnamefont {Smits}}, \bibinfo {author} {\bibfnamefont
  {J.}~\bibnamefont {Damron}}, \bibinfo {author} {\bibfnamefont
  {N.}~\bibnamefont {Ristoff}}, \bibinfo {author} {\bibfnamefont
  {N.}~\bibnamefont {Mosavian}}, \ and\ \bibinfo {author} {\bibfnamefont
  {V.~M.}\ \bibnamefont {Acosta}},\ }\href {\doibase
  10.1103/PhysRevResearch.2.023394} {\bibfield  {journal} {\bibinfo  {journal}
  {Physical Review Research}\ }\textbf {\bibinfo {volume} {2}},\ \bibinfo
  {pages} {23394} (\bibinfo {year} {2020})}\BibitemShut {NoStop}%
\bibitem [{\citenamefont {Glenn}\ \emph {et~al.}(2018)\citenamefont {Glenn},
  \citenamefont {Bucher}, \citenamefont {Lee}, \citenamefont {Lukin},
  \citenamefont {Park},\ and\ \citenamefont {Walsworth}}]{Glenn2018}%
  \BibitemOpen
  \bibfield  {author} {\bibinfo {author} {\bibfnamefont {D.~R.}\ \bibnamefont
  {Glenn}}, \bibinfo {author} {\bibfnamefont {D.~B.}\ \bibnamefont {Bucher}},
  \bibinfo {author} {\bibfnamefont {J.}~\bibnamefont {Lee}}, \bibinfo {author}
  {\bibfnamefont {M.~D.}\ \bibnamefont {Lukin}}, \bibinfo {author}
  {\bibfnamefont {H.}~\bibnamefont {Park}}, \ and\ \bibinfo {author}
  {\bibfnamefont {R.~L.}\ \bibnamefont {Walsworth}},\ }\href {\doibase
  10.1038/nature25781} {\bibfield  {journal} {\bibinfo  {journal} {Nature}\
  }\textbf {\bibinfo {volume} {555}},\ \bibinfo {pages} {351} (\bibinfo {year}
  {2018})}\BibitemShut {NoStop}%
\bibitem [{\citenamefont {Jaskula}\ \emph {et~al.}(2019)\citenamefont
  {Jaskula}, \citenamefont {Saha}, \citenamefont {Ajoy}, \citenamefont
  {Twitchen}, \citenamefont {Markham},\ and\ \citenamefont
  {Cappellaro}}]{Jaskula2019}%
  \BibitemOpen
  \bibfield  {author} {\bibinfo {author} {\bibfnamefont {J.~C.}\ \bibnamefont
  {Jaskula}}, \bibinfo {author} {\bibfnamefont {K.}~\bibnamefont {Saha}},
  \bibinfo {author} {\bibfnamefont {A.}~\bibnamefont {Ajoy}}, \bibinfo {author}
  {\bibfnamefont {D.~J.}\ \bibnamefont {Twitchen}}, \bibinfo {author}
  {\bibfnamefont {M.}~\bibnamefont {Markham}}, \ and\ \bibinfo {author}
  {\bibfnamefont {P.}~\bibnamefont {Cappellaro}},\ }\href {\doibase
  10.1103/PhysRevApplied.11.054010} {\bibfield  {journal} {\bibinfo  {journal}
  {Physical Review Applied}\ }\textbf {\bibinfo {volume} {11}},\ \bibinfo
  {pages} {1} (\bibinfo {year} {2019})}\BibitemShut {NoStop}%
\bibitem [{\citenamefont {Steinert}\ \emph {et~al.}(2010)\citenamefont
  {Steinert}, \citenamefont {Dolde}, \citenamefont {Neumann}, \citenamefont
  {Aird}, \citenamefont {Naydenov}, \citenamefont {Balasubramanian},
  \citenamefont {Jelezko},\ and\ \citenamefont {Wrachtrup}}]{OMDR_Wrachtrup}%
  \BibitemOpen
  \bibfield  {author} {\bibinfo {author} {\bibfnamefont {S.}~\bibnamefont
  {Steinert}}, \bibinfo {author} {\bibfnamefont {F.}~\bibnamefont {Dolde}},
  \bibinfo {author} {\bibfnamefont {P.}~\bibnamefont {Neumann}}, \bibinfo
  {author} {\bibfnamefont {A.}~\bibnamefont {Aird}}, \bibinfo {author}
  {\bibfnamefont {B.}~\bibnamefont {Naydenov}}, \bibinfo {author}
  {\bibfnamefont {G.}~\bibnamefont {Balasubramanian}}, \bibinfo {author}
  {\bibfnamefont {F.}~\bibnamefont {Jelezko}}, \ and\ \bibinfo {author}
  {\bibfnamefont {J.}~\bibnamefont {Wrachtrup}},\ }\href {\doibase
  10.1063/1.3385689} {\bibfield  {journal} {\bibinfo  {journal} {Review of
  Scientific Instruments}\ }\textbf {\bibinfo {volume} {81}},\ \bibinfo {pages}
  {043705} (\bibinfo {year} {2010})}\BibitemShut {NoStop}%
\bibitem [{Hal(2019)}]{HallSensor1}%
  \BibitemOpen
  \href@noop {} {\emph {\bibinfo {title} {A1304 Linear Hall-Effect Sensor IC
  with Analog Output, Available in a Miniature, Low-Profile Surface Mount
  Package}}},\ \bibinfo {organization} {Allegro microsystems} (\bibinfo {year}
  {2019}),\ \bibinfo {note} {rev. 3}\BibitemShut {NoStop}%
\bibitem [{Hal(2020)}]{HallSensor2}%
  \BibitemOpen
  \href@noop {} {\emph {\bibinfo {title} {A1324, A1325, and A1326 Low-Noise
  Linear Hall-Effect Sensor ICs with Analog Output}}},\ \bibinfo {organization}
  {Allegro microsystems} (\bibinfo {year} {2020}),\ \bibinfo {note} {rev.
  8}\BibitemShut {NoStop}%
\bibitem [{\citenamefont {Ramsden}(2011)}]{HallSensor3}%
  \BibitemOpen
  \bibfield  {author} {\bibinfo {author} {\bibfnamefont {E.}~\bibnamefont
  {Ramsden}},\ }\href@noop {} {\emph {\bibinfo {title} {Hall-Effect
  Sensors}}},\ \bibinfo {edition} {2nd}\ ed.\ (\bibinfo  {publisher} {Newnes},\
  \bibinfo {year} {2011})\ \bibinfo {note} {theory and Application}\BibitemShut
  {NoStop}%
\bibitem [{\citenamefont {Grosz}\ \emph {et~al.}(2017)\citenamefont {Grosz},
  \citenamefont {J.~Haji-Sheikh},\ and\ \citenamefont
  {Mukhopadhyay}}]{bookmag}%
  \BibitemOpen
  \bibfield  {author} {\bibinfo {author} {\bibfnamefont {A.}~\bibnamefont
  {Grosz}}, \bibinfo {author} {\bibfnamefont {M.}~\bibnamefont
  {J.~Haji-Sheikh}}, \ and\ \bibinfo {author} {\bibfnamefont {S.~C.}\
  \bibnamefont {Mukhopadhyay}},\ }\href@noop {} {\emph {\bibinfo {title} {High
  Sensitivity Magnetometers}}},\ Vol.~\bibinfo {volume} {19}\ (\bibinfo
  {publisher} {Springer},\ \bibinfo {year} {2017})\BibitemShut {NoStop}%
\bibitem [{\citenamefont {{Dufay}}\ \emph {et~al.}(2013)\citenamefont
  {{Dufay}}, \citenamefont {{Saez}}, \citenamefont {{Dolabdjian}},
  \citenamefont {{Yelon}},\ and\ \citenamefont {{Menard}}}]{GMI}%
  \BibitemOpen
  \bibfield  {author} {\bibinfo {author} {\bibfnamefont {B.}~\bibnamefont
  {{Dufay}}}, \bibinfo {author} {\bibfnamefont {S.}~\bibnamefont {{Saez}}},
  \bibinfo {author} {\bibfnamefont {C.}~\bibnamefont {{Dolabdjian}}}, \bibinfo
  {author} {\bibfnamefont {A.}~\bibnamefont {{Yelon}}}, \ and\ \bibinfo
  {author} {\bibfnamefont {D.}~\bibnamefont {{Menard}}},\ }\href@noop {}
  {\bibfield  {journal} {\bibinfo  {journal} {IEEE Transactions on Magnetics}\
  }\textbf {\bibinfo {volume} {49}},\ \bibinfo {pages} {85} (\bibinfo {year}
  {2013})}\BibitemShut {NoStop}%
\bibitem [{\citenamefont {Kirtley}\ \emph {et~al.}(1995)\citenamefont
  {Kirtley}, \citenamefont {Ketchen}, \citenamefont {Stawiasz}, \citenamefont
  {Sun}, \citenamefont {Gallagher}, \citenamefont {Blanton},\ and\
  \citenamefont {Wind}}]{Kirtley1995}%
  \BibitemOpen
  \bibfield  {author} {\bibinfo {author} {\bibfnamefont {J.~R.}\ \bibnamefont
  {Kirtley}}, \bibinfo {author} {\bibfnamefont {M.~B.}\ \bibnamefont
  {Ketchen}}, \bibinfo {author} {\bibfnamefont {K.~G.}\ \bibnamefont
  {Stawiasz}}, \bibinfo {author} {\bibfnamefont {J.~Z.}\ \bibnamefont {Sun}},
  \bibinfo {author} {\bibfnamefont {W.~J.}\ \bibnamefont {Gallagher}}, \bibinfo
  {author} {\bibfnamefont {S.~H.}\ \bibnamefont {Blanton}}, \ and\ \bibinfo
  {author} {\bibfnamefont {S.~J.}\ \bibnamefont {Wind}},\ }\href {\doibase
  10.1063/1.113838} {\bibfield  {journal} {\bibinfo  {journal} {Applied Physics
  Letters}\ }\textbf {\bibinfo {volume} {66}},\ \bibinfo {pages} {1138}
  (\bibinfo {year} {1995})}\BibitemShut {NoStop}%
\bibitem [{\citenamefont {Finkler}\ \emph {et~al.}(2010)\citenamefont
  {Finkler}, \citenamefont {Segev}, \citenamefont {Myasoedov}, \citenamefont
  {Rappaport}, \citenamefont {Ne'Eman}, \citenamefont {Vasyukov}, \citenamefont
  {Zeldov}, \citenamefont {Huber}, \citenamefont {Martin},\ and\ \citenamefont
  {Yacoby}}]{Finkler2010}%
  \BibitemOpen
  \bibfield  {author} {\bibinfo {author} {\bibfnamefont {A.}~\bibnamefont
  {Finkler}}, \bibinfo {author} {\bibfnamefont {Y.}~\bibnamefont {Segev}},
  \bibinfo {author} {\bibfnamefont {Y.}~\bibnamefont {Myasoedov}}, \bibinfo
  {author} {\bibfnamefont {M.~L.}\ \bibnamefont {Rappaport}}, \bibinfo {author}
  {\bibfnamefont {L.}~\bibnamefont {Ne'Eman}}, \bibinfo {author} {\bibfnamefont
  {D.}~\bibnamefont {Vasyukov}}, \bibinfo {author} {\bibfnamefont
  {E.}~\bibnamefont {Zeldov}}, \bibinfo {author} {\bibfnamefont {M.~E.}\
  \bibnamefont {Huber}}, \bibinfo {author} {\bibfnamefont {J.}~\bibnamefont
  {Martin}}, \ and\ \bibinfo {author} {\bibfnamefont {A.}~\bibnamefont
  {Yacoby}},\ }\href {\doibase 10.1021/nl100009r} {\bibfield  {journal}
  {\bibinfo  {journal} {Nano Letters}\ }\textbf {\bibinfo {volume} {10}},\
  \bibinfo {pages} {1046} (\bibinfo {year} {2010})}\BibitemShut {NoStop}%
\bibitem [{\citenamefont {Nagendran}\ \emph {et~al.}(2011)\citenamefont
  {Nagendran}, \citenamefont {Thirumurugan}, \citenamefont {Chinnasamy},
  \citenamefont {Janawadkar},\ and\ \citenamefont {Sundar}}]{HighFieldSQUID}%
  \BibitemOpen
  \bibfield  {author} {\bibinfo {author} {\bibfnamefont {R.}~\bibnamefont
  {Nagendran}}, \bibinfo {author} {\bibfnamefont {N.}~\bibnamefont
  {Thirumurugan}}, \bibinfo {author} {\bibfnamefont {N.}~\bibnamefont
  {Chinnasamy}}, \bibinfo {author} {\bibfnamefont {M.~P.}\ \bibnamefont
  {Janawadkar}}, \ and\ \bibinfo {author} {\bibfnamefont {C.~S.}\ \bibnamefont
  {Sundar}},\ }\href {\doibase 10.1063/1.3519017} {\bibfield  {journal}
  {\bibinfo  {journal} {Review of Scientific Instruments}\ }\textbf {\bibinfo
  {volume} {82}},\ \bibinfo {pages} {015109} (\bibinfo {year}
  {2011})}\BibitemShut {NoStop}%
\bibitem [{\citenamefont {Stepanov}\ \emph {et~al.}(2015)\citenamefont
  {Stepanov}, \citenamefont {Cho}, \citenamefont {Abeywardana},\ and\
  \citenamefont {Takahashi}}]{Stepanov2015}%
  \BibitemOpen
  \bibfield  {author} {\bibinfo {author} {\bibfnamefont {V.}~\bibnamefont
  {Stepanov}}, \bibinfo {author} {\bibfnamefont {F.~H.}\ \bibnamefont {Cho}},
  \bibinfo {author} {\bibfnamefont {C.}~\bibnamefont {Abeywardana}}, \ and\
  \bibinfo {author} {\bibfnamefont {S.}~\bibnamefont {Takahashi}},\ }\href
  {\doibase 10.1063/1.4908528} {\bibfield  {journal} {\bibinfo  {journal}
  {Applied Physics Letters}\ }\textbf {\bibinfo {volume} {106}},\ \bibinfo
  {pages} {063111} (\bibinfo {year} {2015})}\BibitemShut {NoStop}%
\bibitem [{\citenamefont {Wrachtrup}\ \emph {et~al.}(1993)\citenamefont
  {Wrachtrup}, \citenamefont {von Borczyskowski}, \citenamefont {Bernard},
  \citenamefont {Orrit},\ and\ \citenamefont {Brown}}]{ODMR_Wrachtrup}%
  \BibitemOpen
  \bibfield  {author} {\bibinfo {author} {\bibfnamefont {J.}~\bibnamefont
  {Wrachtrup}}, \bibinfo {author} {\bibfnamefont {C.}~\bibnamefont {von
  Borczyskowski}}, \bibinfo {author} {\bibfnamefont {J.}~\bibnamefont
  {Bernard}}, \bibinfo {author} {\bibfnamefont {M.}~\bibnamefont {Orrit}}, \
  and\ \bibinfo {author} {\bibfnamefont {R.}~\bibnamefont {Brown}},\ }\href
  {\doibase 10.1038/363244a0} {\bibfield  {journal} {\bibinfo  {journal}
  {Nature}\ }\textbf {\bibinfo {volume} {363}},\ \bibinfo {pages} {244}
  (\bibinfo {year} {1993})}\BibitemShut {NoStop}%
\bibitem [{\citenamefont {Iv\'ady}\ \emph {et~al.}(2014)\citenamefont
  {Iv\'ady}, \citenamefont {Simon}, \citenamefont {Maze}, \citenamefont
  {Abrikosov},\ and\ \citenamefont {Gali}}]{Temp_dep}%
  \BibitemOpen
  \bibfield  {author} {\bibinfo {author} {\bibfnamefont {V.}~\bibnamefont
  {Iv\'ady}}, \bibinfo {author} {\bibfnamefont {T.}~\bibnamefont {Simon}},
  \bibinfo {author} {\bibfnamefont {J.~R.}\ \bibnamefont {Maze}}, \bibinfo
  {author} {\bibfnamefont {I.~A.}\ \bibnamefont {Abrikosov}}, \ and\ \bibinfo
  {author} {\bibfnamefont {A.}~\bibnamefont {Gali}},\ }\href {\doibase
  10.1103/PhysRevB.90.235205} {\bibfield  {journal} {\bibinfo  {journal} {Phys.
  Rev. B}\ }\textbf {\bibinfo {volume} {90}},\ \bibinfo {pages} {235205}
  (\bibinfo {year} {2014})}\BibitemShut {NoStop}%
\bibitem [{\citenamefont {Welford}\ and\ \citenamefont {Winston}(1978)}]{CPC}%
  \BibitemOpen
  \bibfield  {author} {\bibinfo {author} {\bibfnamefont {W.}~\bibnamefont
  {Welford}}\ and\ \bibinfo {author} {\bibfnamefont {R.}~\bibnamefont
  {Winston}},\ }\href@noop {} {\emph {\bibinfo {title} {The Optics of
  Nonimaging Concentrators: Light and Solar Energy}}}\ (\bibinfo  {publisher}
  {Academic Press, New York},\ \bibinfo {year} {1978})\BibitemShut {NoStop}%
\bibitem [{\citenamefont {Hobbs}(1992)}]{Autobal_Patent}%
  \BibitemOpen
  \bibfield  {author} {\bibinfo {author} {\bibfnamefont {P.~C.~D.}\
  \bibnamefont {Hobbs}},\ }\href
  {https://patents.google.com/patent/US5134276A/en} {\enquote {\bibinfo {title}
  {Noise cancelling circuitry for optical systems with signal dividing and
  combining means},}\ } (\bibinfo {year} {1992}),\ \bibinfo {note} {~US Patent
  5,134,276}\BibitemShut {NoStop}%
\bibitem [{\citenamefont {Hobbs}(2008)}]{Autobal_Book}%
  \BibitemOpen
  \bibfield  {author} {\bibinfo {author} {\bibfnamefont {P.~C.~D.}\
  \bibnamefont {Hobbs}},\ }\href {\doibase 10.1002/9780470466339.fmatter}
  {\emph {\bibinfo {title} {Building Electro‐Optical Systems: Making it all
  work}}}\ (\bibinfo  {publisher} {John Wiley \& Sons, Ltd},\ \bibinfo {year}
  {2008})\BibitemShut {NoStop}%
\bibitem [{\citenamefont {{Neuhaus}}\ \emph {et~al.}(2017)\citenamefont
  {{Neuhaus}}, \citenamefont {{Metzdorff}}, \citenamefont {{Chua}},
  \citenamefont {{Jacqmin}}, \citenamefont {{Briant}}, \citenamefont
  {{Heidmann}}, \citenamefont {{Cohadon}},\ and\ \citenamefont
  {{Deléglise}}}]{PyRPL_paper}%
  \BibitemOpen
  \bibfield  {author} {\bibinfo {author} {\bibfnamefont {L.}~\bibnamefont
  {{Neuhaus}}}, \bibinfo {author} {\bibfnamefont {R.}~\bibnamefont
  {{Metzdorff}}}, \bibinfo {author} {\bibfnamefont {S.}~\bibnamefont {{Chua}}},
  \bibinfo {author} {\bibfnamefont {T.}~\bibnamefont {{Jacqmin}}}, \bibinfo
  {author} {\bibfnamefont {T.}~\bibnamefont {{Briant}}}, \bibinfo {author}
  {\bibfnamefont {A.}~\bibnamefont {{Heidmann}}}, \bibinfo {author}
  {\bibfnamefont {P.~.}\ \bibnamefont {{Cohadon}}}, \ and\ \bibinfo {author}
  {\bibfnamefont {S.}~\bibnamefont {{Deléglise}}},\ }in\ \href {\doibase
  10.1109/CLEOE-EQEC.2017.8087380} {\emph {\bibinfo {booktitle} {2017
  Conference on Lasers and Electro-Optics Europe European Quantum Electronics
  Conference (CLEO/Europe-EQEC)}}}\ (\bibinfo {year} {2017})\ pp.\ \bibinfo
  {pages} {1--1}\BibitemShut {NoStop}%
\bibitem [{\citenamefont {Vijayan}\ \emph {et~al.}(2020)\citenamefont
  {Vijayan}, \citenamefont {Sompet}, \citenamefont {Salomon}, \citenamefont
  {Koepsell}, \citenamefont {Hirthe}, \citenamefont {Bohrdt}, \citenamefont
  {Grusdt}, \citenamefont {Bloch},\ and\ \citenamefont
  {Gross}}]{LongAveraging}%
  \BibitemOpen
  \bibfield  {author} {\bibinfo {author} {\bibfnamefont {J.}~\bibnamefont
  {Vijayan}}, \bibinfo {author} {\bibfnamefont {P.}~\bibnamefont {Sompet}},
  \bibinfo {author} {\bibfnamefont {G.}~\bibnamefont {Salomon}}, \bibinfo
  {author} {\bibfnamefont {J.}~\bibnamefont {Koepsell}}, \bibinfo {author}
  {\bibfnamefont {S.}~\bibnamefont {Hirthe}}, \bibinfo {author} {\bibfnamefont
  {A.}~\bibnamefont {Bohrdt}}, \bibinfo {author} {\bibfnamefont
  {F.}~\bibnamefont {Grusdt}}, \bibinfo {author} {\bibfnamefont
  {I.}~\bibnamefont {Bloch}}, \ and\ \bibinfo {author} {\bibfnamefont
  {C.}~\bibnamefont {Gross}},\ }\href {\doibase 10.1126/science.aay2354}
  {\bibfield  {journal} {\bibinfo  {journal} {Science}\ }\textbf {\bibinfo
  {volume} {367}},\ \bibinfo {pages} {186} (\bibinfo {year}
  {2020})}\BibitemShut {NoStop}%
\bibitem [{\citenamefont {Schweigler}\ \emph {et~al.}(2017)\citenamefont
  {Schweigler}, \citenamefont {Kasper}, \citenamefont {Erne}, \citenamefont
  {Mazets}, \citenamefont {Rauer}, \citenamefont {Cataldini}, \citenamefont
  {Langen}, \citenamefont {Gasenzer}, \citenamefont {Berges},\ and\
  \citenamefont {Schmiedmayer}}]{LongAveragingSchmiedmayer}%
  \BibitemOpen
  \bibfield  {author} {\bibinfo {author} {\bibfnamefont {T.}~\bibnamefont
  {Schweigler}}, \bibinfo {author} {\bibfnamefont {V.}~\bibnamefont {Kasper}},
  \bibinfo {author} {\bibfnamefont {S.}~\bibnamefont {Erne}}, \bibinfo {author}
  {\bibfnamefont {I.}~\bibnamefont {Mazets}}, \bibinfo {author} {\bibfnamefont
  {B.}~\bibnamefont {Rauer}}, \bibinfo {author} {\bibfnamefont
  {F.}~\bibnamefont {Cataldini}}, \bibinfo {author} {\bibfnamefont
  {T.}~\bibnamefont {Langen}}, \bibinfo {author} {\bibfnamefont
  {T.}~\bibnamefont {Gasenzer}}, \bibinfo {author} {\bibfnamefont
  {J.}~\bibnamefont {Berges}}, \ and\ \bibinfo {author} {\bibfnamefont
  {J.}~\bibnamefont {Schmiedmayer}},\ }\href {\doibase 10.1038/nature22310}
  {\bibfield  {journal} {\bibinfo  {journal} {Nature}\ }\textbf {\bibinfo
  {volume} {545}},\ \bibinfo {pages} {323} (\bibinfo {year}
  {2017})}\BibitemShut {NoStop}%
\bibitem [{\citenamefont {Clevenson}\ \emph {et~al.}(2018)\citenamefont
  {Clevenson}, \citenamefont {Pham}, \citenamefont {Teale}, \citenamefont
  {Johnson}, \citenamefont {Englund},\ and\ \citenamefont
  {Braje}}]{Clevenson2}%
  \BibitemOpen
  \bibfield  {author} {\bibinfo {author} {\bibfnamefont {H.}~\bibnamefont
  {Clevenson}}, \bibinfo {author} {\bibfnamefont {L.~M.}\ \bibnamefont {Pham}},
  \bibinfo {author} {\bibfnamefont {C.}~\bibnamefont {Teale}}, \bibinfo
  {author} {\bibfnamefont {K.}~\bibnamefont {Johnson}}, \bibinfo {author}
  {\bibfnamefont {D.}~\bibnamefont {Englund}}, \ and\ \bibinfo {author}
  {\bibfnamefont {D.}~\bibnamefont {Braje}},\ }\href {\doibase
  10.1063/1.5034216} {\bibfield  {journal} {\bibinfo  {journal} {Applied
  Physics Letters}\ }\textbf {\bibinfo {volume} {112}},\ \bibinfo {pages}
  {252406} (\bibinfo {year} {2018})}\BibitemShut {NoStop}%
\bibitem [{\citenamefont {Eisenach}\ \emph {et~al.}(2020)\citenamefont
  {Eisenach}, \citenamefont {Barry}, \citenamefont {O'Keeffe}, \citenamefont
  {Schloss}, \citenamefont {Steinecker}, \citenamefont {Englund},\ and\
  \citenamefont {Braje}}]{Eisenach2020}%
  \BibitemOpen
  \bibfield  {author} {\bibinfo {author} {\bibfnamefont {E.~R.}\ \bibnamefont
  {Eisenach}}, \bibinfo {author} {\bibfnamefont {J.~F.}\ \bibnamefont {Barry}},
  \bibinfo {author} {\bibfnamefont {M.~F.}\ \bibnamefont {O'Keeffe}}, \bibinfo
  {author} {\bibfnamefont {J.~M.}\ \bibnamefont {Schloss}}, \bibinfo {author}
  {\bibfnamefont {M.~H.}\ \bibnamefont {Steinecker}}, \bibinfo {author}
  {\bibfnamefont {D.~R.}\ \bibnamefont {Englund}}, \ and\ \bibinfo {author}
  {\bibfnamefont {D.~A.}\ \bibnamefont {Braje}},\ }\href@noop {} {\  (\bibinfo
  {year} {2020})},\ \Eprint {http://arxiv.org/abs/2003.01104} {2003.01104}
  \BibitemShut {NoStop}%
\bibitem [{\citenamefont {Ebel}\ \emph {et~al.}(2020)\citenamefont {Ebel},
  \citenamefont {Joas}, \citenamefont {Schalk}, \citenamefont {Angerer},
  \citenamefont {Majer},\ and\ \citenamefont {Reinhard}}]{Ebel2020}%
  \BibitemOpen
  \bibfield  {author} {\bibinfo {author} {\bibfnamefont {J.}~\bibnamefont
  {Ebel}}, \bibinfo {author} {\bibfnamefont {T.}~\bibnamefont {Joas}}, \bibinfo
  {author} {\bibfnamefont {M.}~\bibnamefont {Schalk}}, \bibinfo {author}
  {\bibfnamefont {A.}~\bibnamefont {Angerer}}, \bibinfo {author} {\bibfnamefont
  {J.}~\bibnamefont {Majer}}, \ and\ \bibinfo {author} {\bibfnamefont
  {F.}~\bibnamefont {Reinhard}},\ }\href@noop {} {\  (\bibinfo {year}
  {2020})}\BibitemShut {NoStop}%
\bibitem [{\citenamefont {Blakley}\ \emph {et~al.}(2015)\citenamefont
  {Blakley}, \citenamefont {Fedotov}, \citenamefont {Kilin},\ and\
  \citenamefont {Zheltikov}}]{NVfiber}%
  \BibitemOpen
  \bibfield  {author} {\bibinfo {author} {\bibfnamefont {S.~M.}\ \bibnamefont
  {Blakley}}, \bibinfo {author} {\bibfnamefont {I.~V.}\ \bibnamefont
  {Fedotov}}, \bibinfo {author} {\bibfnamefont {S.~Y.}\ \bibnamefont {Kilin}},
  \ and\ \bibinfo {author} {\bibfnamefont {A.~M.}\ \bibnamefont {Zheltikov}},\
  }\href {\doibase 10.1364/OL.40.003727} {\bibfield  {journal} {\bibinfo
  {journal} {Opt. Lett.}\ }\textbf {\bibinfo {volume} {40}},\ \bibinfo {pages}
  {3727} (\bibinfo {year} {2015})}\BibitemShut {NoStop}%
\bibitem [{\citenamefont {Fedotov}\ \emph
  {et~al.}(2014{\natexlab{a}})\citenamefont {Fedotov}, \citenamefont
  {Doronina-Amitonova}, \citenamefont {Sidorov-Biryukov}, \citenamefont
  {Safronov}, \citenamefont {Blakley}, \citenamefont {Levchenko}, \citenamefont
  {Zibrov}, \citenamefont {Fedotov}, \citenamefont {Kilin}, \citenamefont
  {Scully}, \citenamefont {Velichansky},\ and\ \citenamefont
  {Zheltikov}}]{NVfiber2}%
  \BibitemOpen
  \bibfield  {author} {\bibinfo {author} {\bibfnamefont {I.~V.}\ \bibnamefont
  {Fedotov}}, \bibinfo {author} {\bibfnamefont {L.~V.}\ \bibnamefont
  {Doronina-Amitonova}}, \bibinfo {author} {\bibfnamefont {D.~A.}\ \bibnamefont
  {Sidorov-Biryukov}}, \bibinfo {author} {\bibfnamefont {N.~A.}\ \bibnamefont
  {Safronov}}, \bibinfo {author} {\bibfnamefont {S.}~\bibnamefont {Blakley}},
  \bibinfo {author} {\bibfnamefont {A.~O.}\ \bibnamefont {Levchenko}}, \bibinfo
  {author} {\bibfnamefont {S.~A.}\ \bibnamefont {Zibrov}}, \bibinfo {author}
  {\bibfnamefont {A.~B.}\ \bibnamefont {Fedotov}}, \bibinfo {author}
  {\bibfnamefont {S.~Y.}\ \bibnamefont {Kilin}}, \bibinfo {author}
  {\bibfnamefont {M.~O.}\ \bibnamefont {Scully}}, \bibinfo {author}
  {\bibfnamefont {V.~L.}\ \bibnamefont {Velichansky}}, \ and\ \bibinfo {author}
  {\bibfnamefont {A.~M.}\ \bibnamefont {Zheltikov}},\ }\href {\doibase
  10.1364/OL.39.006954} {\bibfield  {journal} {\bibinfo  {journal} {Opt.
  Lett.}\ }\textbf {\bibinfo {volume} {39}},\ \bibinfo {pages} {6954} (\bibinfo
  {year} {2014}{\natexlab{a}})}\BibitemShut {NoStop}%
\bibitem [{\citenamefont {Fedotov}\ \emph
  {et~al.}(2014{\natexlab{b}})\citenamefont {Fedotov}, \citenamefont
  {Doronina-Amitonova}, \citenamefont {Voronin}, \citenamefont {Levchenko},
  \citenamefont {Zibrov}, \citenamefont {Sidorov-Biryukov}, \citenamefont
  {Fedotov}, \citenamefont {Velichansky},\ and\ \citenamefont
  {Zheltikov}}]{Fedotov2014}%
  \BibitemOpen
  \bibfield  {author} {\bibinfo {author} {\bibfnamefont {I.~V.}\ \bibnamefont
  {Fedotov}}, \bibinfo {author} {\bibfnamefont {L.~V.}\ \bibnamefont
  {Doronina-Amitonova}}, \bibinfo {author} {\bibfnamefont {A.~A.}\ \bibnamefont
  {Voronin}}, \bibinfo {author} {\bibfnamefont {A.~O.}\ \bibnamefont
  {Levchenko}}, \bibinfo {author} {\bibfnamefont {S.~A.}\ \bibnamefont
  {Zibrov}}, \bibinfo {author} {\bibfnamefont {D.~A.}\ \bibnamefont
  {Sidorov-Biryukov}}, \bibinfo {author} {\bibfnamefont {A.~B.}\ \bibnamefont
  {Fedotov}}, \bibinfo {author} {\bibfnamefont {V.~L.}\ \bibnamefont
  {Velichansky}}, \ and\ \bibinfo {author} {\bibfnamefont {A.~M.}\ \bibnamefont
  {Zheltikov}},\ }\href {\doibase 10.1038/srep05362} {\bibfield  {journal}
  {\bibinfo  {journal} {Scientific Reports}\ }\textbf {\bibinfo {volume} {4}},\
  \bibinfo {pages} {5362} (\bibinfo {year} {2014}{\natexlab{b}})}\BibitemShut
  {NoStop}%
\bibitem [{\citenamefont {Chomaz}\ \emph {et~al.}(2019)\citenamefont {Chomaz},
  \citenamefont {Petter}, \citenamefont {Ilzh\"ofer}, \citenamefont {Natale},
  \citenamefont {Trautmann}, \citenamefont {Politi}, \citenamefont
  {Durastante}, \citenamefont {van Bijnen}, \citenamefont {Patscheider},
  \citenamefont {Sohmen}, \citenamefont {Mark},\ and\ \citenamefont
  {Ferlaino}}]{SupersolidFerl}%
  \BibitemOpen
  \bibfield  {author} {\bibinfo {author} {\bibfnamefont {L.}~\bibnamefont
  {Chomaz}}, \bibinfo {author} {\bibfnamefont {D.}~\bibnamefont {Petter}},
  \bibinfo {author} {\bibfnamefont {P.}~\bibnamefont {Ilzh\"ofer}}, \bibinfo
  {author} {\bibfnamefont {G.}~\bibnamefont {Natale}}, \bibinfo {author}
  {\bibfnamefont {A.}~\bibnamefont {Trautmann}}, \bibinfo {author}
  {\bibfnamefont {C.}~\bibnamefont {Politi}}, \bibinfo {author} {\bibfnamefont
  {G.}~\bibnamefont {Durastante}}, \bibinfo {author} {\bibfnamefont {R.~M.~W.}\
  \bibnamefont {van Bijnen}}, \bibinfo {author} {\bibfnamefont
  {A.}~\bibnamefont {Patscheider}}, \bibinfo {author} {\bibfnamefont
  {M.}~\bibnamefont {Sohmen}}, \bibinfo {author} {\bibfnamefont {M.~J.}\
  \bibnamefont {Mark}}, \ and\ \bibinfo {author} {\bibfnamefont
  {F.}~\bibnamefont {Ferlaino}},\ }\href {\doibase 10.1103/PhysRevX.9.021012}
  {\bibfield  {journal} {\bibinfo  {journal} {Phys. Rev. X}\ }\textbf {\bibinfo
  {volume} {9}},\ \bibinfo {pages} {021012} (\bibinfo {year}
  {2019})}\BibitemShut {NoStop}%
\bibitem [{\citenamefont {B\"ottcher}\ \emph {et~al.}(2019)\citenamefont
  {B\"ottcher}, \citenamefont {Schmidt}, \citenamefont {Wenzel}, \citenamefont
  {Hertkorn}, \citenamefont {Guo}, \citenamefont {Langen},\ and\ \citenamefont
  {Pfau}}]{SupersolidPfau}%
  \BibitemOpen
  \bibfield  {author} {\bibinfo {author} {\bibfnamefont {F.}~\bibnamefont
  {B\"ottcher}}, \bibinfo {author} {\bibfnamefont {J.-N.}\ \bibnamefont
  {Schmidt}}, \bibinfo {author} {\bibfnamefont {M.}~\bibnamefont {Wenzel}},
  \bibinfo {author} {\bibfnamefont {J.}~\bibnamefont {Hertkorn}}, \bibinfo
  {author} {\bibfnamefont {M.}~\bibnamefont {Guo}}, \bibinfo {author}
  {\bibfnamefont {T.}~\bibnamefont {Langen}}, \ and\ \bibinfo {author}
  {\bibfnamefont {T.}~\bibnamefont {Pfau}},\ }\href {\doibase
  10.1103/PhysRevX.9.011051} {\bibfield  {journal} {\bibinfo  {journal} {Phys.
  Rev. X}\ }\textbf {\bibinfo {volume} {9}},\ \bibinfo {pages} {011051}
  (\bibinfo {year} {2019})}\BibitemShut {NoStop}%
\bibitem [{\citenamefont {Li}\ \emph {et~al.}(2015)\citenamefont {Li},
  \citenamefont {Zhu}, \citenamefont {He}, \citenamefont {Wang}, \citenamefont
  {Guo}, \citenamefont {Xu}, \citenamefont {Zhang},\ and\ \citenamefont
  {Wang}}]{CohHeteroSpin}%
  \BibitemOpen
  \bibfield  {author} {\bibinfo {author} {\bibfnamefont {X.}~\bibnamefont
  {Li}}, \bibinfo {author} {\bibfnamefont {B.}~\bibnamefont {Zhu}}, \bibinfo
  {author} {\bibfnamefont {X.}~\bibnamefont {He}}, \bibinfo {author}
  {\bibfnamefont {F.}~\bibnamefont {Wang}}, \bibinfo {author} {\bibfnamefont
  {M.}~\bibnamefont {Guo}}, \bibinfo {author} {\bibfnamefont {Z.-F.}\
  \bibnamefont {Xu}}, \bibinfo {author} {\bibfnamefont {S.}~\bibnamefont
  {Zhang}}, \ and\ \bibinfo {author} {\bibfnamefont {D.}~\bibnamefont {Wang}},\
  }\href {\doibase 10.1103/PhysRevLett.114.255301} {\bibfield  {journal}
  {\bibinfo  {journal} {Phys. Rev. Lett.}\ }\textbf {\bibinfo {volume} {114}},\
  \bibinfo {pages} {255301} (\bibinfo {year} {2015})}\BibitemShut {NoStop}%
\bibitem [{\citenamefont {Loubser}\ and\ \citenamefont {{Van
  Wyk}}(1978)}]{Loubser1978}%
  \BibitemOpen
  \bibfield  {author} {\bibinfo {author} {\bibfnamefont {J.~H.}\ \bibnamefont
  {Loubser}}\ and\ \bibinfo {author} {\bibfnamefont {J.~A.}\ \bibnamefont {{Van
  Wyk}}},\ }\href {\doibase 10.1088/0034-4885/41/8/002} {\bibfield  {journal}
  {\bibinfo  {journal} {Reports on Progress in Physics}\ }\textbf {\bibinfo
  {volume} {41}},\ \bibinfo {pages} {1201} (\bibinfo {year}
  {1978})}\BibitemShut {NoStop}%
\bibitem [{\citenamefont {Doherty}\ \emph {et~al.}(2011)\citenamefont
  {Doherty}, \citenamefont {Manson}, \citenamefont {Delaney},\ and\
  \citenamefont {Hollenberg}}]{Doherty2011}%
  \BibitemOpen
  \bibfield  {author} {\bibinfo {author} {\bibfnamefont {M.~W.}\ \bibnamefont
  {Doherty}}, \bibinfo {author} {\bibfnamefont {N.~B.}\ \bibnamefont {Manson}},
  \bibinfo {author} {\bibfnamefont {P.}~\bibnamefont {Delaney}}, \ and\
  \bibinfo {author} {\bibfnamefont {L.~C.}\ \bibnamefont {Hollenberg}},\ }\href
  {\doibase 10.1088/1367-2630/13/2/025019} {\bibfield  {journal} {\bibinfo
  {journal} {New Journal of Physics}\ }\textbf {\bibinfo {volume} {13}},\
  \bibinfo {pages} {025019} (\bibinfo {year} {2011})}\BibitemShut {NoStop}%
\bibitem [{\citenamefont {Zhukov}\ \emph {et~al.}(2017)\citenamefont {Zhukov},
  \citenamefont {Anishchik},\ and\ \citenamefont {Molin}}]{Zhukov2017}%
  \BibitemOpen
  \bibfield  {author} {\bibinfo {author} {\bibfnamefont {I.~V.}\ \bibnamefont
  {Zhukov}}, \bibinfo {author} {\bibfnamefont {S.~V.}\ \bibnamefont
  {Anishchik}}, \ and\ \bibinfo {author} {\bibfnamefont {Y.~N.}\ \bibnamefont
  {Molin}},\ }\href {\doibase 10.1007/s00723-017-0933-6} {\bibfield  {journal}
  {\bibinfo  {journal} {Applied Magnetic Resonance}\ }\textbf {\bibinfo
  {volume} {48}},\ \bibinfo {pages} {1461} (\bibinfo {year}
  {2017})}\BibitemShut {NoStop}%
\bibitem [{\citenamefont {Hesse}\ \emph {et~al.}(2020)\citenamefont {Hesse},
  \citenamefont {Köster}, \citenamefont {Steiner}, \citenamefont {Michl},
  \citenamefont {Vorobyov}, \citenamefont {Dasari}, \citenamefont {Wrachtrup},\
  and\ \citenamefont {Jendrzejewski}}]{Heidata}%
  \BibitemOpen
  \bibfield  {author} {\bibinfo {author} {\bibfnamefont {A.}~\bibnamefont
  {Hesse}}, \bibinfo {author} {\bibfnamefont {K.}~\bibnamefont {Köster}},
  \bibinfo {author} {\bibfnamefont {J.}~\bibnamefont {Steiner}}, \bibinfo
  {author} {\bibfnamefont {J.}~\bibnamefont {Michl}}, \bibinfo {author}
  {\bibfnamefont {V.}~\bibnamefont {Vorobyov}}, \bibinfo {author}
  {\bibfnamefont {D.}~\bibnamefont {Dasari}}, \bibinfo {author} {\bibfnamefont
  {J.}~\bibnamefont {Wrachtrup}}, \ and\ \bibinfo {author} {\bibfnamefont
  {F.}~\bibnamefont {Jendrzejewski}},\ }\href {\doibase 10.11588/data/W60KUN}
  {\enquote {\bibinfo {title} {{Direct control of high magnetic fields for cold
  atom experiments based on NV centers [Dataset]}},}\ } (\bibinfo {year}
  {2020})\BibitemShut {NoStop}%
\bibitem [{\citenamefont {Chubar}\ \emph {et~al.}(1998)\citenamefont {Chubar},
  \citenamefont {Elleaume},\ and\ \citenamefont {Chavanne}}]{Radia}%
  \BibitemOpen
  \bibfield  {author} {\bibinfo {author} {\bibfnamefont {O.}~\bibnamefont
  {Chubar}}, \bibinfo {author} {\bibfnamefont {P.}~\bibnamefont {Elleaume}}, \
  and\ \bibinfo {author} {\bibfnamefont {J.}~\bibnamefont {Chavanne}},\ }\href
  {\doibase 10.1107/S0909049597013502} {\bibfield  {journal} {\bibinfo
  {journal} {Journal of Synchrotron Radiation}\ }\textbf {\bibinfo {volume}
  {5}},\ \bibinfo {pages} {481} (\bibinfo {year} {1998})}\BibitemShut {NoStop}%
\end{thebibliography}%

\end{document}